\documentclass{aa}  

\usepackage{graphicx}
\usepackage{mathtools}
\usepackage{txfonts}
\usepackage{multirow}
\usepackage{caption, threeparttable}
\usepackage{xcolor}
\begin{document}

%%%%%%%%%%%%%%%%%%%%%%%%%%%%%%%%%%%%%%%%%%%%%%%%%%
\newcommand{\teff}{$T_\mathrm{eff}$}
\newcommand{\logg}{$\log g$}
\newcommand{\feh}{[Fe/H]}
\newcommand{\sife}{[Si/Fe]}
\newcommand{\mgfe}{[Mg/Fe]}
\newcommand{\afe}{[$\alpha$/Fe]}
\newcommand{\co}{CO($\nu =7-4$)\,}

\newcommand{\water}{H$_2$O}
\newcommand{\invcm}{cm$^{-1}$}
\newcommand{\kms}{km\,s$^{-1}$}
\newcommand{\mic}{$\mu \mathrm m$}

%%%%%%%%%%%%%%%%%%% TITLE PAGE %%%%%%%%%%%%%%%%%%%

   % \title{Stellar parameters and $\alpha$ abundance trends of M giants in the solar neighborhood with IGRINS}
   \title{ M giants with IGRINS \\
   I. Stellar parameters and $\alpha$-abundance trends of the solar neighborhood population}

   \subtitle{}

   \author{G. Nandakumar
            \inst{1}
            \and
          N. Ryde
          \inst{1}
          \and
          L. Casagrande
            \inst{2,3}
            \and
        G. Mace
        \inst{4,5}
          }

   \institute{Lund Observatory, Division of Astrophysics, Department of Physics, Lund University, Box 43, SE-221 00 Lund, Sweden\\
              \email{govind.nandakumar@fysik.lu.se} 
              \and
             Research School of Astronomy and Astrophysics, The Australian National University, Canberra, ACT 2611, Australia
             \and
            ARC Centre of Excellence for All Sky Astrophysics in 3 Dimensions (ASTRO 3D), Australia 
              \and
    Department of Astronomy, The University of Texas, Austin, TX 78712, USA
    \and 
    McDonald Observatory, The University of Texas, Austin, TX 78712, USA
     }

   \date{Received ; accepted }

% \abstract{}{}{}{}{} 
% 5 {} token are mandatory
 
  \abstract
   { Cool stars, such as M giants, can only be analysed in the near-infrared (NIR) regime due to the ubiquitous TiO features in optical spectra of stars with \teff\, $<$ 4000\,K.  In dust obscured regions, like the inner bulge and Galactic Center region, the intrinsically bright M giants observed in the near-infrared is an optimal option to study stellar abundances and the chemical evolution of stellar populations. Due to uncertainties in photometric methods, a method to determine the stellar parameters for M giants from the near-IR spectra themselves is needed.   }
   { To develop a method to determine the stellar parameters for M giants from the near-IR spectra. Validate the method by deriving the stellar parameters for nearby well-studied M-giants with spectra from the IGRINS spectral library. Demonstrate the accuracy and precision by determining stellar parameters and $\alpha$-element trends versus metallicity for solar neighbourhood M giants.   }
   {We have carried out new observations of 44 M giant stars with IGRINS mounted on the Gemini South telescope within the programs GS-2020B-Q-305 and GS-2021A-Q302. We also obtained the full H and K band IGRINS spectra of six nearby well-studied M giants at a spectral resolving power of R=45,000 from the IGRINS spectral library. We use the Spectroscopy Made Easy (SME) tool in combination with one-dimensional (1D) Model Atmospheres in a Radiative and Convective Scheme (MARCS) stellar atmosphere models to model the synthetic spectrum that best fits the observed spectrum.  }
   { The effective temperatures that we derive from our new method (here tested for 3400$\lesssim$ \teff $\lesssim$4000\,K) agree excellently with six nearby, well-studied M giants which indicates that the accuracy is indeed high. For the 43 solar neighborhood M giants, our \teff, \logg, \feh,  $\xi_\mathrm{micro}$, [C/Fe], [N/Fe], and [O/Fe] are in unison with APOGEE with mean differences and scatter (our method - APOGEE) of -67$\pm$33 K, -0.31$\pm$0.15 dex, 0.02$\pm$0.05 dex, 0.22$\pm$0.13 km/s, -0.05$\pm$0.06 dex, 0.06$\pm$0.06 dex, and 0.02$\pm$0.09 dex, respectively. Furthermore, the tight  offset with a small dispersion compared to APOGEE's \teff\ points to a high precision in both our derived temperatures and those derived from the APOGEE pipeline. Typical uncertainties in the stellar parameters are found to be $\pm$ 100 K in \teff\,, $\pm$0.2 dex in \logg\,, $\pm$0.1 dex in \feh\,, $\pm$0.1 km/s in $\xi_\mathrm{micro}$. The $\alpha$-element trends versus metallicity for Mg, Si, Ca and Ti are consistent with both APOGEE DR17 trends for the same stars as well as with the GILD optical trends. We also find clear enhancement in abundances for thick disc stars. }% contrary to what we expect
   {}


%
%-------------------------------------------------------------------
\titlerunning{M Giants observed with IGRINS}

\authorrunning{ Nandakumar et al.}

\maketitle
%--------------------------------------------------------------------
\section{Introduction}
\label{sec:intro}
As necessary ingredients in any spectroscopic analyses and specifically in order to accurately estimate detailed elemental abundances from stellar spectra, the fundamental parameters of stars are crucial to determine. These are the effective temperature (\teff), surface gravity (\logg), metallicity (\feh) and microturbulence ($\xi_\mathrm{micro}$; in the case of 1D stellar atmospheres). With the advent of bigger telescopes and major advancement in the instruments to efficiently record spectra in the visual or optical wavelength regime, there has been a huge progress in the spectroscopic analysis techniques to estimate accurate fundamental stellar parameters %from optical spectra 
\citep[see, e.g.,][]{jofre:2019}. This has also led to advances in large-scale analyses of optical spectroscopic surveys like GALAH \citep{Buder:2021}, LAMOST \citep{lamost}, Gaia-ESO \citep{GES} and several others which observe and analyse the stellar spectra of millions of stars of different spectral types and at various evolutionary stages. Yet these surveys provide reliable stellar parameters and abundances mainly for the relatively warmer FGK type stars \citep[\teff\, $>$ 4000 K;][]{jofre:2019} since the optical spectra of the ubiquitous, cooler M-type stars are riddled by diatomic (e.g., TiO, FeH, OH, CO) and triatomic (e.g., H2O) molecules in their atmospheres making their optical spectra nearly impossible to analyse. 

Spectra of M-type stars in the near-infrared wavelength regime (0.75 $\mu$m - 2.4 $\mu$m) are, however, feasible to analyse. This opens up the possibility to use these stars in  spectroscopic analyses and to use them as probes for the study of stellar populations.  
%which, even though not completely devoid of molecules, is comparatively less affected by them. 
In addition, stars in direction of high extinction are preferably analysed in the infrared. For instance, spectroscopic investigations of stars in the heavily dust enshrouded Galactic Centre region demand observations at near-infrared wavelengths and also limit the observable stellar populations to the relatively brighter M-giant stars. 
%But as mentioned earlier, the . 
This calls for the development of spectroscopic techniques to determine reliable stellar parameters also from spectra recorded at the near-IR wavelengths. With the immense possibility of observing more distant M giants with the upcoming large telescopes like ELT \citep{ELT} and TMT \citep{TMT} with the high-resolution, near-infrared spectroscopic instruments projected for them \citep{hires:13,hires:16,tmt_nir:19}, there is an urgent need to develop methods to derive reliable stellar parameters from near infrared spectra of M giants.

%  like excitation balance for \teff and ionisation balance for \logg
Existing methods to determine the effective temperatures, \teff, of M giants like those based on interferometry \citep{Mozurkewich:2003}, the infrared flux method \citep[IRFM;][]{casagrande:08}, and the color-\teff\, relation \citep{Bessell:1998} are limited to nearby stars and also subjected to uncertainties arising due to interstellar reddening. Thus, the determination of \teff\, using the above mentioned methods is less useful for the distant and heavily extinct stars in the inner Milky Way regions thus demanding methods that can extract stellar parameters directly from near-infrared spectra. %Spectroscopic \teff\,  determinations from near infrared spectra 
Such methods for GK stars (\teff\, $>$ 4000 K) using the ratios of 
the depths between low- and high-excitation lines (i.e. line depth ratios; \citealt{Gray:2008}) have been attempted in recent studies (\citealt{Fukue:2015}, \citealt{Taniguchi:2018}, \citealt{Taniguchi:2021}, \citealt{Matsunaga:2021}, \citealt{Afsar:2023}) but have been found to be affected by metallicity and abundance ratios \citep{Jian:2019}. \citet{ryde_schultheis:15} and \citet{schultheis:16}  determined \teff\ for M giant stars in the Solar neighbourhood and the  Galactic Centre by using the spectral indices of CO band heads in low-resolution K-band spectra and \citet{thorsbro:18,Thorsbro:2020} used spectral indices of K band Scandium lines. Similarly, \citet{Ghosh:2019,Ghosh:2021,Ghosh:2022} determined metallicity dependent \teff-equivalent width relations using low resolution H- and K- band spectra of cool stars. But these methods need further refinement and tests using reliable benchmark stars. 

Similarly asteroseismology, considered to be one of the reliable methods to determine the surface gravities, \logg,\, of giants, is currently tested only for warmer giants \citep{Pinsonneault:2018}  and also needs high quality repeated observations. Another way is to use the fundamental relation of \logg\, \citep{Nissen:1997} that demands reliable values of mass, luminosity, stellar radius and distance measurements:
\begin{equation}
   \mathrm{log \frac{g}{g_{\odot}}} = log \frac{M}{M_{\odot}} - 4 log \frac{T_\mathrm{eff}}{T_\mathrm{eff_\odot}} + 0.4(M_{bol}-M_{bol_\odot})
\end{equation}
where M is the stellar mass, \teff\, is the effective temperature and M$_{bol}$ is the absolute bolometric magnitude estimated from the distance modulus relation.  
 This again is possible only for bright nearby stars with reliable parallax measurements from astrometry mission like the Gaia mission \citep{gaia_dr3}. This necessitates the development of other methods to determine surface gravity from the near infrared spectra of distant M giants.

In this paper, we present an iterative method using Spectroscopy made easy \citep[SME;][]{sme,sme_code} to determine reliable stellar parameters of cool M giants  ($3400\lesssim$\teff $\lesssim4000$\,K) in the solar neighbourhood from  near-infrared, H-band spectra. We demonstrate the method by using IGRINS spectra at spectral resolving power of $R\sim45,000$ and make use of \teff\, sensitive molecular OH lines in combination with CO molecular bands and Fe lines. The details of the observations and data reduction procedure are described in the section~\ref{sec:obs}. In section~\ref{sec:analysis}, we give a brief description of the line list used in this work and in section~\ref{sec:parameters} we describe our iterative method used to determine stellar parameters for six nearby well studied M giants based on spectra from the  IGRINS spectral library \citep{Park:2018,rrisa} in section~\ref{sec:results}. We show the quality of our parameter estimates by comparing the determined \teff\, with reliable values in literature (from interferometry, photometric and spectroscopic methods) and if available, \logg\, and \feh. With this method, we then determine stellar parameters and $\alpha$-element abundances of 44 solar neighbourhood M giants which are also observed by the near infrared spectroscopic survey, APOGEE \citep[Apache Point Observatory Galactic Evolution Experiment;][]{Holtzman:2018,Jonsson:2018}. In section~\ref{sec:results}, we also compare stellar parameters and the $\alpha$-element abundance trends with respect to APOGEE and other literature sources. Final conclusions are in the section~\ref{sec:conclusion}.

\begin{table*}
\caption{ Observational details of K giant stars. }\label{table:obs}
\centering
\begin{tabular}{c c c c c c c c}
\hline
\hline
 Index & 2MASS ID  & H$_\mathrm{2MASS}$  & K$_\mathrm{2MASS}$  & Date    & Exposure Time & SNR$_\mathrm{H}$ & SNR$_\mathrm{K}$ \\

 &  &  (mag) & (mag) &  UT &  mm:ss  &  \multicolumn{2}{c}{(per resolution element)} \\
\hline
1 & 2M05484106-0602007  & 7.0 & 6.7 &  2021-01-02  &  06:32   & 110  &  130  \\
2 & 2M05594446-7212111  & 7.0 & 6.8 &  2021-01-02  &  10:35   & 100  &  120   \\
3 & 2M06035110-7456029  & 7.0 & 6.7 &  2021-01-01  &  13:25   & 90  &  110  \\
4 & 2M06035214-7255079  & 7.1 & 6.8 &  2021-01-01  &  06:04   & 100  &  110  \\
5 & 2M06052796-0553384  & 7.0 & 6.6 &  2021-01-02  &  07:08   & 100  &  120   \\
6 & 2M06074096-0530332  & 7.3 & 6.5 &  2021-01-02  &  08:22   & 90  &  130   \\
7 & 2M06124201-0025095  & 7.0 & 6.7 &  2021-01-01  &  07:34   & 90  &  100   \\
8 & 2M06140107-0641072  & 7.2 & 6.8 &  2021-01-02  &  06:02   & 50  &  65  \\
9 & 2M06143705-0551064  & 7.2 & 6.9 &  2021-01-02  &  07:13   & 70  &  90   \\
10 & 2M06171159-7259319  & 7.2 & 6.9 &  2021-01-02  &  06:21   & 120  &  130  \\
11 & 2M06223443-0443153  & 7.1 & 6.8 &  2021-01-02  &  08:35   & 90  &  110   \\
12 & 2M06231693-0530385  & 7.1 & 6.7 &  2021-01-02  &  08:05   & 80  &  100   \\
13 & 2M06520463-0047080  & 7.1 & 6.7 &  2021-01-02  &  06:41   & 80  &  100   \\
14 & 2M06551808-0148080  & 7.1 & 6.8 &  2021-01-02  &  04:10   & 90  &  100   \\
15 & 2M06574070-1231239  & 7.3 & 6.9 &  2021-01-02  &  09:31   & 90  &  110   \\
16 & 2M10430394-4605354  & 7.2 & 6.9 &  2021-02-11  &  06:22   & 120  &  120  \\
17 & 2M11042542-7318068  & 7.2 & 6.9 &  2021-02-11  &  05:50   & 120  &  130   \\
18 & 2M12101600-4936072  & 7.2 & 6.9 &  2021-02-19  &  06:28   & 70  &  90    \\
19 & 2M13403516-5040261  & 7.1  & 6.8  &  2021-03-17  &  11:27   & 170  &  170   \\
20 & 2M14131192-4849280  & 7.2  & 6.9  &  2021-03-17  &  09:02   & 150  &  150   \\
21 & 2M14240039-6252516  & 6.6  & 6.2  &  2021-03-18  &  12:36   & 240  &  250  \\
22 & 2M14241044-6218367  & 7.3  & 6.8  &  2021-04-17  &  09:38   & 230  &  260   \\
%-"-                 &  " & "  &  2021-04-20  &  05:55 &  &  \\
23 & 2M14260433-6219024  & 7.2 & 6.8  &  2021-02-18  & 05:01   & 50  &  70  \\
24 & 2M14261117-6240220  & 7.0 & 6.5  &  2021-02-14  & 08:09   & 80  &  100  \\
25 & 2M14275833-6147534  & 7.0 & 6.5  &  2021-02-18  & 05:19   & 80  &  100   \\
26 & 2M14283733-6257279  & 6.5 & 6.1  &  2021-02-18  & 06:59   & 70  &  90  \\
27 & 2M14291063-6317181  & 7.4 & 6.9  &  2021-02-18  & 05:16   & 80  &  110   \\
28 & 2M14311520-6145468  & 7.3 & 6.7  &  2021-04-20  &  06:32  & 150  &  170   \\
%2M14311520-6145468  & 7.3 & 6.7  &  2021-04-17  & 21:52   & 150  &  170   \\
%-"-                 &  " & "  &  2021-04-20  &  06:32 &  &  \\
29 & 2M14322072-6215506  & 6.7 & 6.4  &  2021-04-13  &  11:40 & 270  &  290  \\
%2M14322072-6215506  & 6.7 & 6.4  &  2021-03-04  &  17:40  & 270  &  290  \\
%-"-                 &  " & "  &  2021-04-13  &  11:40 &  &  \\
30 & 2M14332169-6302108  & 6.9 & 6.5  &  2021-04-20  &  05:45  & 120  &  130   \\
31 & 2M14332869-6211255  & 6.5 & 6.1  &  2021-02-20  &  06:20  & 120  &  140  \\
32 & 2M14333081-6221450  & 7.2 & 6.7  &  2021-03-18  &  10:10  & 210  &  240  \\
33 & 2M14333688-6232028  & 7.2 & 6.8  &  2021-04-24  &  14:24  & 210  &  250  \\
34 & 2M14345114-6225509  & 7.0 & 6.6  &  2021-04-24  &  13:55  & 270  &  300  \\
35 & 2M14360142-6228561  & 6.4 & 6.0  &  2021-04-20  &  10:23  & 80 &  90   \\
36 & 2M14360935-6309399  & 6.9 & 6.5  &  2021-04-20  &  06:40  & 90  &  100  \\
37 & 2M14371958-6251344  & 6.7 & 6.3  &  2021-04-24  &  09:18  & 240  &  240   \\
38 & 2M14375085-6237526  & 7.3 & 7.0  &  2021-04-20  &  07:49  & 100  &  120  \\
39 & 2M15161949+0244516  & 7.2 & 7.0  &  2021-04-24  &  14:01  & 190  &  190  \\
40 & 2M17584888-2351011  & 7.3 & 6.5  &  2021-04-28  &  08:21  & 200  &  290  \\
41 & 2M18103303-1626220  & 7.3 & 6.5  &  2021-04-28  &  08:47  & 230  &  310  \\
42 & 2M18142346-2136410  & 7.1 & 6.6  &  2021-04-28  &  12:19  & 240  &  280  \\
43 & 2M18191551-1726223  & 7.2 & 6.6  &  2021-04-28  &  14:16  & 320  &  380  \\
44 & 2M18522108-3022143  & 7.3 & 6.9  &  2021-04-28  &  24:03  & 320  &  320  \\
\hline
\hline                                 %inserts single line
\end{tabular}
\end{table*}

\section{Observations and Data Reduction}
\label{sec:obs}

In this work, we have analysed near-infrared spectra of 50 M giants observed with the Immersion GRating INfrared Spectrograph \citep[IGRINS;][]{Yuk:2010,Wang:2010,Gully:2012,Moon:2012,Park:2014,Jeong:2014}.
IGRINS provides spectra spanning the full H and K bands (1.45 - 2.5 $\mu$m) with a spectral resolving power of $R \sim$ 45,000. We have carried out new observations of 44 M giant stars with IGRINS mounted on the Gemini South telescope \citep{Mace:2018} within the programs GS-2020B-Q-305 and GS-2021A-Q302. The observations were performed in Service Mode in Jan to April 2021.  Details of the observations are listed in the Table~\ref{table:obs}. We also analyse the IGRINS spectra of %use the high resolution ($R \sim$ 100,000) infrared spectrum of Arcturus from the Arcturus atlas \citep{Hinkle:1995} as well as 
six nearby M giants available in the IGRINS spectral library \citep{Park:2018,rrisa}, which we present in Table \ref{table:standard}.
 
The IGRINS observations were carried out in one or more ABBA nod sequences along the slit permitting sky background subtraction. The exposure times were set to aim at an average signal-to-noise ratios (SNR)\footnote{SNR is provided by RRISA \citep[The Raw $\&$ Reduced IGRINS Spectral Archive;][]{rrisa} and is the average SNR for the H or K band and is per resolution element. It varies over the orders and is lowest at the end of the orders} of at least 100, leading to observing times ranging from 5 to 25 minutes. In all cases apart from a few a SNR above 100 was achieved, see last column in Table~\ref{table:obs}. For a third of the observations two or three times the SNR was achieved due better than expected weather conditions. 

We used the IGRINS PipeLine Package \citep[IGRINS PLP;][]{Lee:2017} to optimally extract the telluric corrected,  wavelength calibrated spectra after flat-field correction and A-B frame subtraction. The spectral orders of the science targets and the telluric standards are subsequently stitched together after normalizing every order and then combining them in {\tt iraf} \citep{IRAF}, excluding the low signal-to-noise edges of every order. This resulted in one normalized stitched spectrum for the entire H and K bands. However, to take care of any modulations in the continuum levels of the spectra we put large attention in defining specific local continua around the spectral line being studied. This turns out to be an important measure for accurate determinations of the $\alpha$ element abundances, see for example \citet{pablo:20}. In the subsequent abundance analysis we also allow for wavelength shifts in order to fit the lines in the line masks we will use. This will take care of any errors or trends in the wavelength solution.

The standard procedure for eliminating the contaminating telluric lines is to divide with a telluric standard-star spectrum, showing only telluric lines and mostly no stellar features.  In the IGRINS observing strategy, telluric stars were chosen to be rapidly rotating, late B to early A dwarfs,
and observed close in time and at a similar air mass as the science targets. This procedure works very well for most wavelength regions. Apart from some broad Brackett lines of hydrogen\footnote{{ Brackett (n=7) line at 2166 nm}}, some spurious broadband spectral features might, however, turn up in the telluric standard-star spectrum. Special attention is given to lines in these regions.

% (see \citet{sameshita:18} to understand the limitations of using standard telluric correction by A dwarfs and the method proposed to  t 

 In high-resolution spectra there might also appear broad absorption features due to Diffuse Interstellar Bands \citep[DIBs, see, e.g.,][]{dib:geballe1}. These are probably due to 
large molecules in the Interstellar Medium (ISM) in the line-of-sight of the star in study. The DIBs are often weak and 
normally correlate with the reddening, E$_{\mathrm{(B-V)}}$, thus mainly found in spectra of reddened stars. Most DIBs have been identified in optical spectra but some are found in the Near-IR \citep{dib:geballe_nature}, and more are being identified with new instruments 
\citep[e.g. with X-shooter, APOGEE, IGRINS, WINERED, and CRIRES in][respectively]{dib:cox,dib:elya,dib:galazut,dib:winered,dib_crires}. None of these known DIBs are close to the spectral lines that we use in the following discussion in this paper.

Finally, for the wavelength solutions, sky OH emission lines are used \citep{Han:2012,Oh:2014} and the spectra are subsequently shifted to laboratory wavelengths in air after a stellar radial velocity correction. In addition, we make sure to carefully eliminate obvious cosmic-ray signatures in the spectra. 

\section{Analysis}
\label{sec:analysis}

In this work, spectral synthesis is carried out using the Spectroscopy Made Easy code \citep[SME;][]{sme,sme_code}. SME generates synthetic spectra by calculating the  spherical radiative transfer through a relevant stellar atmosphere model defined by its fundamental, stellar parameters. SME finds this model by interpolating in a grid of  one-dimensional (1D) MARCS (Model Atmospheres in a Radiative and Convective Scheme) stellar atmosphere models \citep{marcs:08}. These are hydrostatic model atmospheres in spherical geometry, computed assuming LTE, chemical equilibrium, homogeneity, and conservation of the total flux. In order to account for the non-LTE (NLTE) effects, we used a NLTE grid for which the departure coefficients were computed using the MPI-parallelised NLTE radiative transfer code \texttt{Balder} \citep{Amarsi:2018}. The NLTE grids for Si, Mg, and Ca are from \cite{NLTE} and for Fe from \cite{lind17} and \cite{amarsi16} (with subsequent updates Amarsi priv. comm.). SME applies departure coefficients by interpolating in these grids.
% In order to determine the stellar parameters directly from the H band spectrum, we have developed an iterative method using SME which will be described in the section below.

%\subsection{Linelist}
\label{sec:linelist}

As a line list we used an updated version of the VALD linelist \citep{vald,vald4,vald5} in this work. The transition probability ($gf$-values, i.e., product of statistical weight and oscillator strength) of the lines lacking reliable experimental $gf$-values have been determined astrophysically using the high resolution infrared solar flux spectrum of \cite{Wallace:2003} and tested for Arcturus ($\alpha$boo) using the high resolution ($R \sim$ 100,000) infrared spectrum from the Arcturus atlas \citep{Hinkle:1995} and adopting the stellar parameters from \cite{Ramirez:2011}. In the astrophysical determination of $gf$-values, we set the abundance of an element to a reference value (0.0 in the case of the Sun and from \citealt{Ramirez:2011} for Arcturus) and fit a synthetic spectrum to the absorption line in the observed spectrum of the Sun or Arcturus by varying the $gf$-value.  

For many lines, we adopted the broadening parameters (corresponding to the collisional broadening due to neutral hydrogen, and in some cases charged particles) from the ABO theory \citep{anstee_investigation_1991,Anstee1995,Barklem1997a,Barklem1998b} or from the spectral synthesis code BSYN based on routines from MARCS \citep{marcs:08}. Details about the calculation of the broadening parameters for the magnesium multiplet lines in the K band can be found in \cite{Nieuwmunster:2023}. 

A detailed validation of the linelist has been carried out using the IGRINS H and K band spectra of $\sim$ 40 solar neighborhood K giants in the Giants in the Local disk (GILD) sample (J\"onsson et al. in prep., which builds upon and improves the analysis described in \citet{jonsson:17}). The stellar parameters and abundances for stars in the GILD sample are determined from optical FIES spectra. The linelist used in this work has been validated by comparing the elemental abundance trends determined from IGRINS near infrared absorption lines for the above mentioned K giants with the trends for the same stars determined in GILD with a reliable optical linelist (Nandakumar et al. in prep). The linelist used in this work have been used in \citealt{montelius:22, Nandakumar:2022, Nieuwmunster:2023}. Thus we use a linelist with carefully selected and reliable lines that have been tested not only with the Sun and the Arcturus but also with $\sim$ 40 solar neighborhood K giants.  

The line data for the CO, CN, and OH molecular lines are adopted from the line lists of \citet{li:2015}, \citet{brooke:2016} and \citet{sneden:2014}, respectively. The central wavelengths and $gf$-values values of the selected OH, CN, and CO molecular lines used in the stellar parameter estimation method (see Section~\ref{sec:parameters}) are listed in the Tables~\ref{table:OH},~\ref{table:CN}, and \ref{table:CO}, respectively. The central wavelengths, VALD or astrophysically calibrated $gf$-values and broadening parameters of Mg, Si, Ca, Ti and Fe are listed in the Table~\ref{table:lines}.

% \hline
% \multirow{12}{*}{Fe} & 20798.893 $^{6}$ & -3.525 $^{6}$ & -7.280 $^{6 }$ &    \\
%  & 20799.685 $^{6}$ & -0.613 $^{6*}$ & -7.021 $^{6*}$ &    \\
%  & 20805.113 $^{7 *}$ & -0.065 $^{7 *}$ & -7.076 $^{7 *}$ &    \\
%  & 20840.835 $^{6 *}$ & 0.104 $^{6 *}$ & -7.134 $^{6 *}$ &    \\
%  & 20882.233 $^{6 }$ & -0.935 $^{6 *}$ & -7.330 $^{6 }$ &    \\
%  & 20948.086 $^{6}$ & -0.796 $^{6 *}$ & -6.861 $^{6 *}$ &    \\
%  & 20991.083 $^{6 *}$ & -3.058 $^{6 *}$ & -7.730 $^{6 }$ &    \\
%  & 21036.355 $^{6}$ & -0.817 $^{6 *}$ & 2656 $^{2}$ &  0.330  \\
%  & 21095.401 $^{6}$ & -0.658 $^{6 *}$ & -7.000 $^{6 *}$ &    \\
%  & 21105.217 $^{6}$ & -0.749 $^{6 *}$ & -7.320 $^{6 }$ &    \\
%  & 21124.505 $^{6 *}$ & -1.647 $^{6 *}$ & 975 $^{2}$ &  0.302  \\
%  & 21162.095 $^{6 *}$ & -0.605 $^{6 *}$ & 1364 $^{2}$ &  0.328  \\

\section{Method}
\label{sec:parameters}

An accurate determination of the effective temperature, \teff, is crucial %in the spectral synthesis techniques employed 
in spectroscopic analyses, as \teff\, defines the energy transported through a star thus shaping of the continuum of the stellar spectrum, the excitation balance of lines, and ionization stages of elements. Hence, we started by identifying \teff-sensitive absorption lines in the H band regime that can be used to constrain \teff. We chose a random set of 236 stars from the APOGEE DR17 catalog within 100 K bins between 3000 K $<$ \teff\, $<$ 4500 K, and 0.5 dex bins between -1.5 $<$ [Fe/H] $<$ 0.5 dex. Based on careful visual inspection of APOGEE spectra of stars with different \teff\, but similar set of \logg, \feh, $\xi_\mathrm{micro}$ etc, we identified a set of $\sim$50 molecular OH-lines that are sensitive to \teff. We further selected a subset of 15 to 20 OH lines from which we were able to recover the APOGEE \teff\, within $\pm$ 100 K when we run SME with \teff\, set as a free parameter for the 236 stars. In addition to \teff, the strength of these lines are also dependent on the oxygen (O) abundance. Thus, it is necessary for this method that the O abundance be known or fixed in order to constrain the \teff\, from OH lines.

\begin{figure}
  \includegraphics[width=\columnwidth]{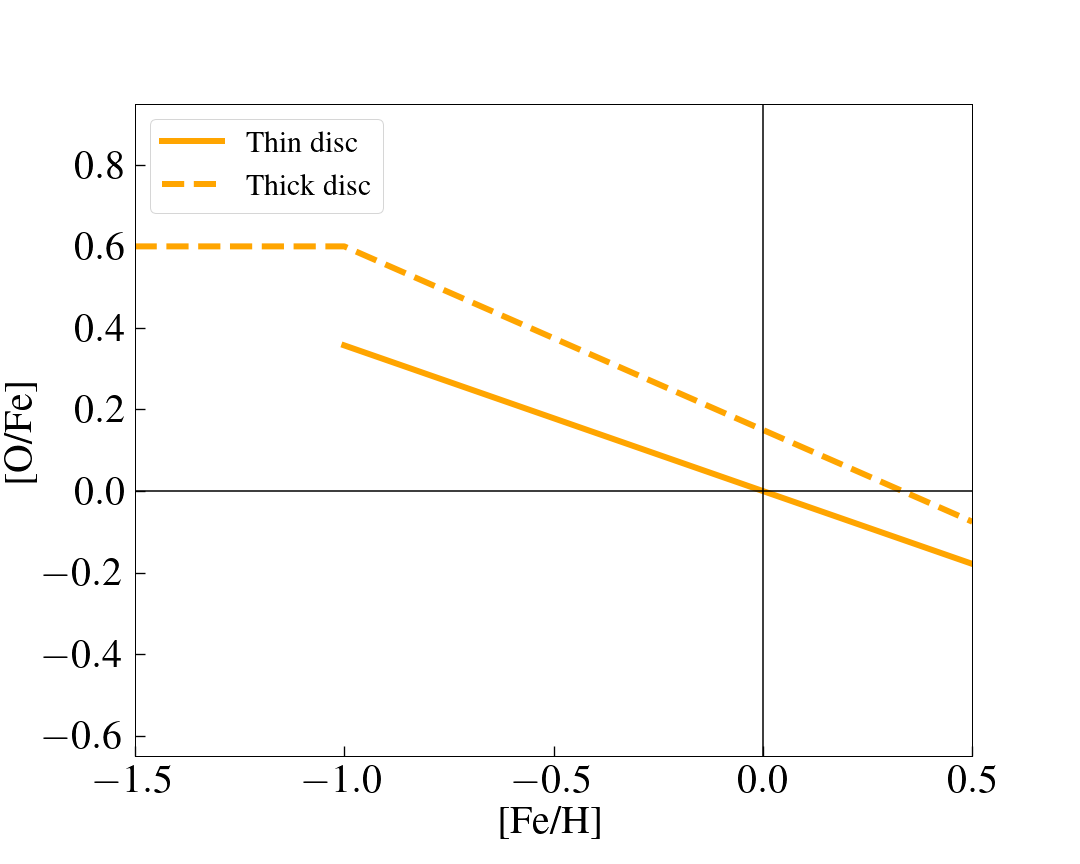}
  \caption{A simple functional form of the [O/Fe] versus [Fe/H] trend for thin and thick disc stellar populations \citep[or better 'low-$\alpha$' and 'high-$\alpha$' populations; see ][]{minchev:17} adopted from \cite{Amarsi:2019}. In our parameter estimation method, we assumed [O/Fe] values for our stars corresponding to a particular metallicity value depending on whether they belong to the thin or thick disc population. }
  \label{fig:Otrend}%
\end{figure}

\begin{figure*}
  \includegraphics[width=\textwidth]{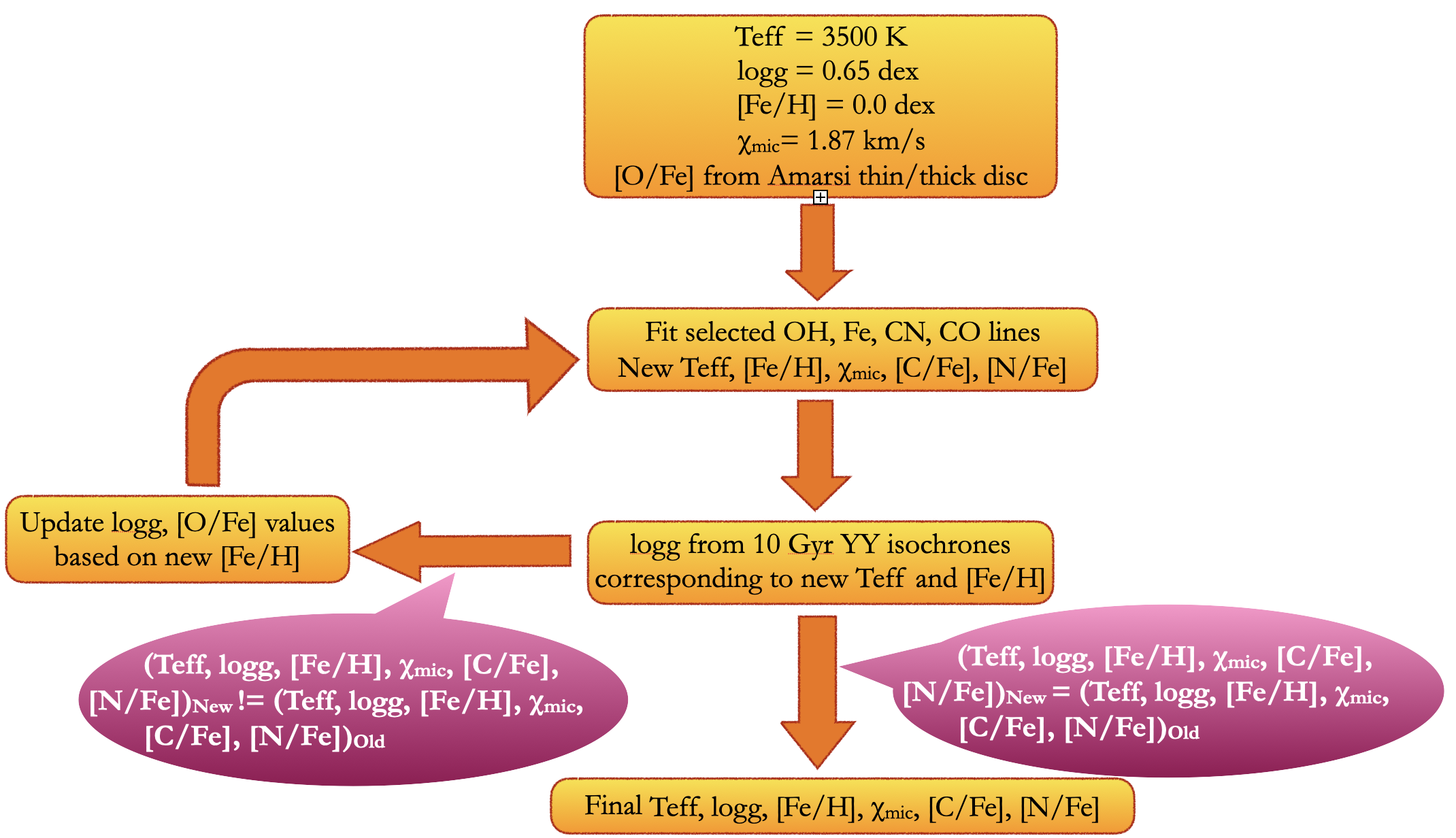}
  \caption{Flow diagram of the method to determine reliable stellar parameters of cool M giants  ($3400\lesssim$\teff $\lesssim4000$\,K)  from  near-infrared, H-band spectra. }
  \label{fig:flow}%
\end{figure*}

\cite{Amarsi:2019} derived 3D NLTE O abundances for 187 F and G dwarfs belonging to thin disc, thick disc and halo of the Milky Way. Based on their Figure 12 (left panel) we have made functional forms of the [O/Fe] versus [Fe/H] trend for thin and thick discs, as shown in Figure~\ref{fig:Otrend}. We used these functional forms of trends to fix [O/Fe] for stars with any metallicity belonging to the thin or the thick disc\footnote{It would actually be more correct to call these populations 'low $\alpha$ population' instead of a thin disc population and 'high $\alpha$ population instead of a thick disc population \citep[see, e.g., ][]{minchev:17}.}. 

In addition to the OH lines, we have chosen a set of CN and CO lines (including molecular band heads) and Fe lines as listed in Tables~\ref{table:CN}, ~\ref{table:CO} and ~\ref{table:lines} respectively. This will constrain the molecular equilibrium of C, N, and O bearing molecules, and the metallicity.

As the first assumption, we categorise the star into a thin disc or thick disc star. We then assume a starting \teff\, and \feh\, of 3500 K and 0.00 dex, respectively. For this \teff\, and \feh\, we get a \logg\, of 0.65 for a 10 Gyr old star from the Yonsei-Yale isochrones by means of simple linear interpolation  \citep[see also][]{rich:17}. At this metallicity, [O/Fe] is 0.0 for a thin disc star and 0.15 for a thick disc star. In the initial step, we run SME setting \teff, \feh,  $\xi_\mathrm{micro}$, C, and N abundances as free parameters for the selected set of lines. SME generates and fits multiple synthetic spectra for all the chosen lines of interest for different combinations of the free parameters. The final values of the parameters are determined from the model with the best fits of the chosen lines by means of $\chi^{2}$ minimization between the synthesized spectrum and the observed one in the marked regions of the chosen spectral lines. This initial step results in a new set of \teff\, and \feh\,, which can be used to constrain \logg\, from the Yonsei-Yale (YY) isochrones assuming old ages of 3-10 Gyr \citep{Demarque:2004}. In the next SME run, we use the \teff\,, \feh\, $\xi_\mathrm{micro}$, C and N abundances from the previous run, \logg\, from the YY isochrone tracks and [O/Fe] at the new \feh\, based on the trend in Figure~\ref{fig:Otrend}. This cycle is repeated until the difference between values of all free parameters from the current SME run and the previous SME run is negligible. In the Figure~\ref{fig:flow} we show the flow diagram indicating the sequence of steps followed in the method.

While \teff\, and \feh\, are mainly constrained by the OH lines and Fe lines respectively, different sets of weak and strong lines help constrain the $\xi_\mathrm{micro}$. Inclusion of CO and CN lines not only constrain the C, N abundances but also results in excellent synthetic spectra fit to the observed CN, CO lines thus taking care of the CN, CO blends as well. At the same time, when we carry out the entire exercise excluding the CN, CO lines and thus removing C and N from the set of free parameters, there is negligible difference in all stellar parameter values except \feh\, which is found to vary within 0.1 dex. This could be an indication of possible CN, and/or CO blends in the Fe lines we have selected. 

% In the H band, aluminium (Al) and titanium (Ti) have been found to be sensitive to \teff\,, at the same time dependent on Al and Ti abundances. Thus in order to determine \teff\, from these atomic lines by spectrum synthesis using SME, their abundances should be known and fixed. Since we

% \begin{table*}
%   \begin{center}
% \caption{The fundamental stellar parameters estimated from the IGRINS spectra of six benchmark stars using our method and the literature compilation of stellar parameters. }\label{table:references}
% \begin{tabular}{c c c c}
%  Star  & \teff  & \logg  & \feh    \\
% \hline
% \hline
% \multirow{3}{*}{HD 132813} & Optical interferometry $^{1}$ & -- & -- \\ 
%  &  \teff-(V-K) based on angular diameter measurements from \cite{Richichi:1999} $^{2}$  & Fundamental relation $^{2}$  & -- \\
%  &  Full spectrum fitting $^{3}$  &  Full spectrum fitting $^{3}$  &  Full spectrum fitting $^{3}$
% \hline
% \hline
% \end{tabular}
% \end{center}
% \tablefoot{\\ 1: \cite{Baines:2021}, 2: \cite{Lebzelter:2019}, 3: \cite{Sharma:2016}, 4: \teff\, from \cite{Bessell:1998} V-K relation, 5: \cite{Jonsson:2014}, 6: \cite{Guerco:2019}, 7: \cite{Smith:1990} }
% \end{table*}

\begin{table*}
  \begin{center}
\caption{The fundamental stellar parameters estimated from the IGRINS spectra of six benchmark stars using our method and the literature compilation of stellar parameters (indicated by the $lit$ subscript). The literature references are listed in the footnote and are indicated by the corresponding number as superscript to the parameter value in the table. We have listed \logg\, and \feh\, values as '--' if there are no \logg, \feh\, measurements in the literature (for example, in case of optical interferometric measurements like in \cite{Baines:2021} or \teff\, versus (V-K)$_{0}$ relation in \cite{Bessell:1988}) }\label{table:standard}
\begin{tabular}{c c c c c c c c c c c}
 Star &  V  &  K$_{2MASS}$  &  E(B-V)  & \teff\, (K) & $T_\mathrm{eff}$ $_{lit}$ (K) & \logg\,  (dex) & $\log g$ $_{lit}$ (dex) & \feh\,  (dex) & [Fe/H] $_{lit}$ (dex)  \\
\hline
\hline
\multirow{4}{*}{HD 132813} & \multirow{4}{*}{4.54} & \multirow{4}{*}{-0.96}  & \multirow{4}{*}{0.02}  & \multirow{4}{*}{3457} & 3410 $\pm$ 37 $^{1}$ & \multirow{4}{*}{0.43}  &  --  & \multirow{4}{*}{-0.27}  &  --  \\
 &  & & &  &    3406 $^{2}$  &  &  0.53 $^{2}$  &  &  --     \\
 &  & & &  &    3387 $\pm$ 39 $^{3}$  &  &  0.46 $\pm$ 0.25 $^{3}$  &  &  -0.09 $\pm$ 0.18 $^{3}$   \\
 &  & & &  &    3458 $^{4}$  &  &  --  &  &  --     \\
\hline
\hline
\multirow{4}{*}{HD 89758} & \multirow{4}{*}{3.05} & \multirow{4}{*}{-1.01} & \multirow{4}{*}{0.01} & \multirow{4}{*}{3807} & 3868 $\pm$ 37 $^{1}$ & \multirow{4}{*}{1.15}  &  --  & \multirow{4}{*}{-0.09}  &  --      \\
 &  & & &  &    3793 $\pm$ 50 $^{5}$  &  &  1.07 $\pm$ 0.10 $^{5}$  &  &  -0.34 $\pm$ 0.10 $^{5}$          \\
 &  & & &  &    3822 $\pm$ 43 $^{3}$  &  &  1.39 $\pm$ 0.17 $^{3}$  &  &  -0.20  $\pm$ 0.06 $^{3}$       \\ 
 &  & & &  &    3777 $^{4}$  &  &  --  &  &  --     \\ 
\hline
\hline
\multirow{4}{*}{HD 175588} & \multirow{4}{*}{4.3} & \multirow{4}{*}{-1.25} & \multirow{4}{*}{0.11} & \multirow{4}{*}{3484} & 3394 $\pm$ 32 $^{1}$ & \multirow{4}{*}{0.49}  &  --  & \multirow{4}{*}{-0.04}  &  --      \\
 &  & & &  &    3408 $^{2}$  &  &  --  &  &  --     \\
 &  & & &  &    3484 $\pm$ 16 $^{3}$  &  &  0.47 $\pm$ 0.17 $^{3}$  &  &  -0.14 $\pm$ 0.16 $^{3}$   \\ 
 &  & & &  &    3487 $^{4}$  &  &  --  &  &  --     \\ 
\hline
\hline
\multirow{3}{*}{HD 224935} & \multirow{4}{*}{4.41} & \multirow{4}{*}{-0.40} & \multirow{4}{*}{0.04} & \multirow{3}{*}{3529} & 3490 $\pm$ 35 $^{1}$ & \multirow{3}{*}{0.64}  &  --  & \multirow{3}{*}{-0.10}  &  --      \\
 &  & & &  &    3504 $^{2}$  &  &  --  &  &  --     \\
 &  & & &  &    3592 $^{4}$  &  &  --  &  &  --     \\
\hline
\hline
\multirow{3}{*}{HD 101153} & \multirow{4}{*}{5.36} & \multirow{4}{*}{-0.21} & \multirow{4}{*}{0.04} & \multirow{3}{*}{3438} & 3421 $\pm$ 35 $^{3}$ & \multirow{3}{*}{0.51}  &  0.48 $\pm$ 0.25 $^{3}$  & \multirow{3}{*}{-0.07}  &  -0.09 $\pm$ 0.12 $^{3}$      \\
 &  & & &  &    3418 $\pm$ 100 $^{6}$  &  &  0.49 $\pm$ 0.25 $^{6}$  &  &  0.00 $\pm$ 0.10 $^{6}$   \\
 &  & & &  &    3455 $^{4}$  &  &  --  &  &  --     \\
\hline
\hline
\multirow{4}{*}{HD 96360} & \multirow{4}{*}{8.09} & \multirow{4}{*}{2.77} & \multirow{4}{*}{0.02} & \multirow{4}{*}{3459} & 3432 $^{2}$ & \multirow{4}{*}{0.5}  &  --  & \multirow{4}{*}{-0.15}  &  --      \\
 &  & & &  &    3471 $\pm$ 22 $^{3}$  &  &  0.8 $\pm$ 0.23 $^{3}$  &  &  0.0 $\pm$ 0.17 $^{3}$   \\ 
 &  & & &  &    3550 $\pm$ 100 $^{6,7}$  &  &  0.72 $\pm$ 0.25 $^{6,7}$  &  &  -0.37 $\pm$ 0.10 $^{6,7}$   \\
 &  & & &  &    3484 $^{4}$  &  &  --  &  &  --     \\
\hline
\hline
\end{tabular}
\end{center}
\tablefoot{\\ 1: \cite{Baines:2021}, 2: \cite{Lebzelter:2019}, 3: \cite{Sharma:2016}, 4: \teff\, from \cite{Bessell:1998} V-K relation, 5: \cite{Jonsson:2014}, 6: \cite{Guerco:2019}, 7: \cite{Smith:1990} }
\end{table*}

\begin{figure*}
  \includegraphics[width=\textwidth]{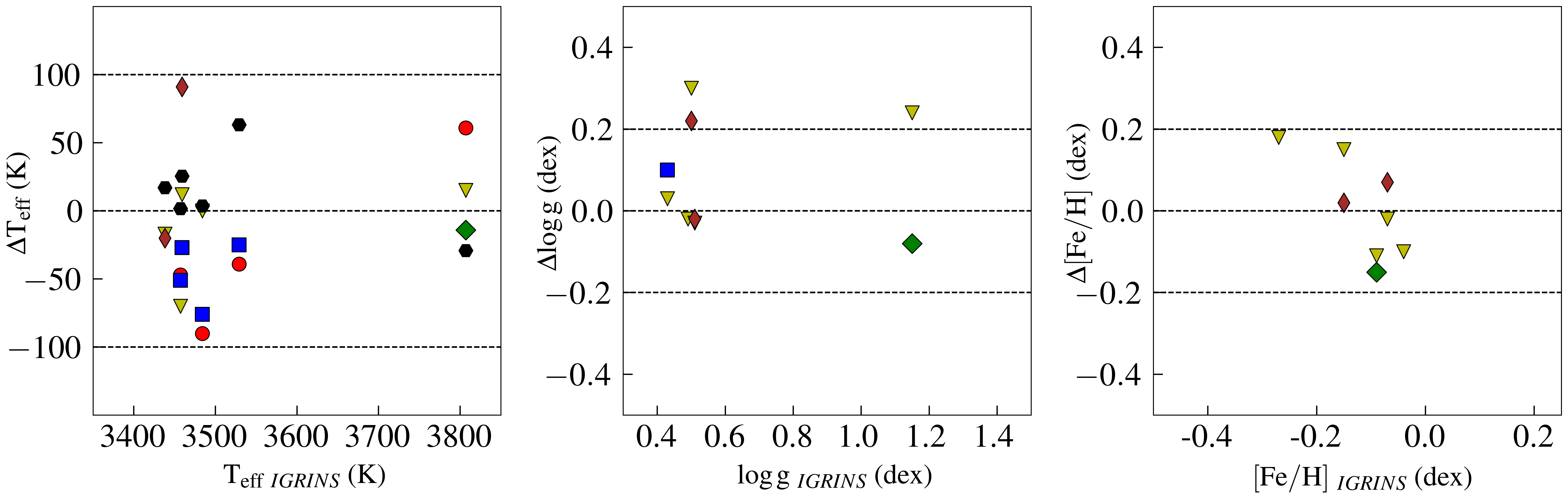}
  \caption{ The difference (literature - this work) in \teff\,(left panel), \logg\,(middle panel), and \feh\,(right panel) on the the y-axis versus the respective parameter estimates using our method (x-axis) for the six nearby M giant stars in the IGRINS spectral library. Different colored symbols represent the compiled literature estimates: red circles - \cite{Baines:2021}, blue squares - \cite{Lebzelter:2019}, green diamonds - \cite{Jonsson:2014}, yellow inverted triangles - \cite{Sharma:2016}, brown diamonds - \cite{Guerco:2019}. The black circles denote the \teff\, estimated using the \teff\, versus (V-K)$_{0}$ relation in Equation~\ref{eqn:bessel} \citep[from ][]{Bessell:1998}. K$_{2MASS}$ has been corrected to the photometric system in \cite{Bessell:1988}  (https://irsa.ipac.caltech.edu/data/2MASS/docs/releases/allsky/doc/sec6$\_$4b.html)   }
  \label{fig:benchmark}%
\end{figure*}

%\section{Results and Discussions}
\section{Validation of the Method}
\label{sec:results}

In this section, we validate our method described in the previous section by determining stellar parameters for six nearby M giants (see Section~\ref{sec:nearbyM}) some of which have reliable parameters in the literature. We then use the method to determine stellar parameters for 44 solar neighborhood M giants (see Section~\ref{sec:SNM}) followed by a discussion on uncertainties (see Section~\ref{sec:uncertain}). We later determine the  $\alpha$ abundance trends from selected Mg, Si, Ca and Ti lines (Section~\ref{sec:alphaabund}) which further proves the usefulness of the method by showing the precision and accuracy of the abundance trends compared to those determined based on other methods. 

\subsection{Stellar parameters}
\label{sec:stellarparams}

\begin{figure*}
  \includegraphics[width=\textwidth]{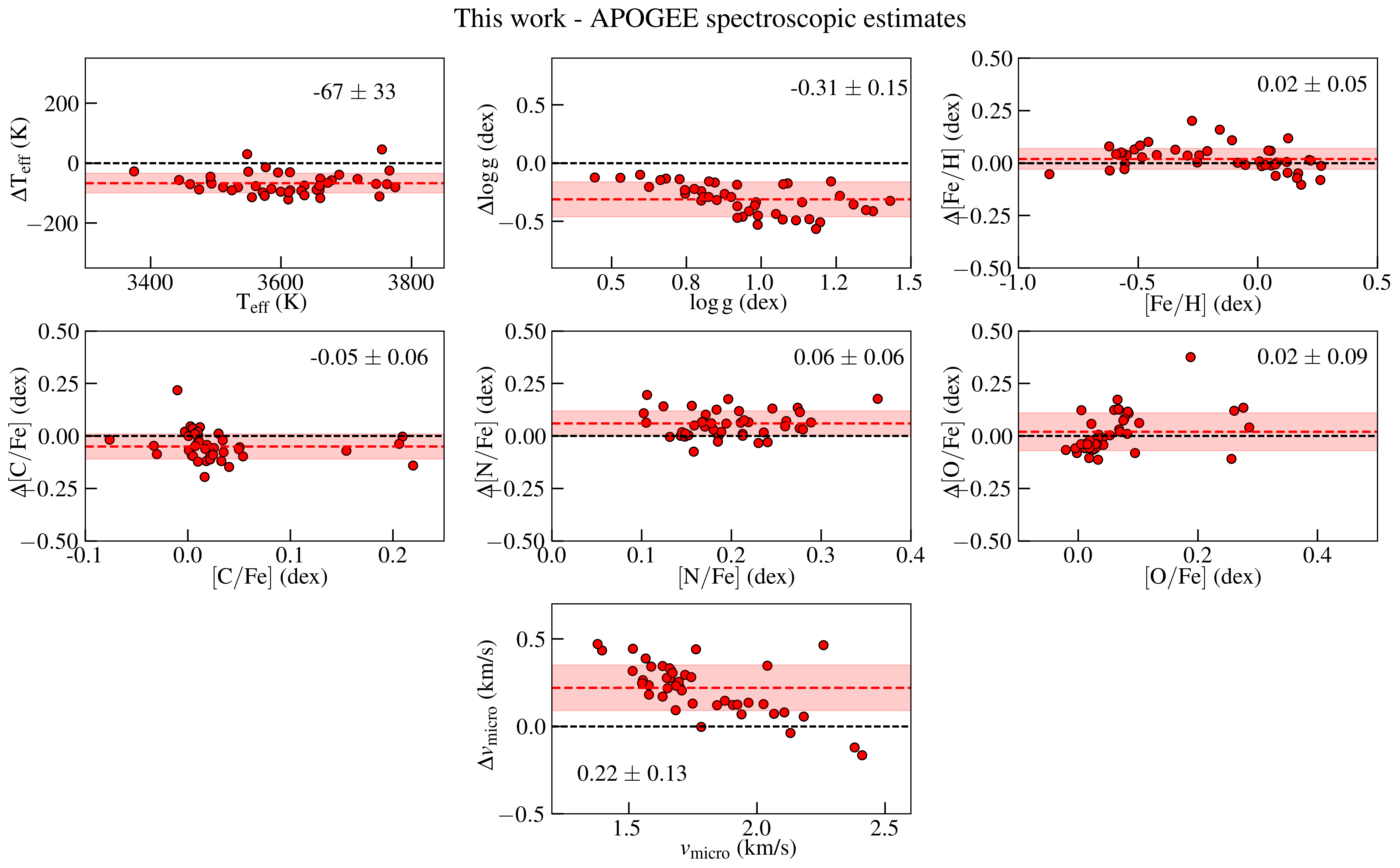}
  \caption{ The differences (this work - APOGEE spectroscopic estimates) in \teff, \logg, \feh,  $\xi_\mathrm{micro}$, [C/Fe], [N/Fe], and [O/Fe] on the y-axis versus the APOGEE spectroscopic estimates on the x-axis for the 44 stars in the solar neighborhood from our new IGRINS observations. \teff, \logg, and \feh are shown in the three panels in the top row, [C/Fe], [N/Fe], and [O/Fe] in the three panels in the middle row, and $\xi_\mathrm{micro}$ in the bottom row panel. The black dashed line indicate difference value of 0.0 between APOGEE and our estimates. The mean difference and standard deviation (calculated as the mid value of 84$^{th}$ - 16$^{th}$ percentile values for each parameter is indicated by the red dashed line and the red band respectively, and is also listed in the respective panels.}
  \label{fig:paramdiff}
\end{figure*}

\begin{figure*}
  \includegraphics[width=\textwidth]{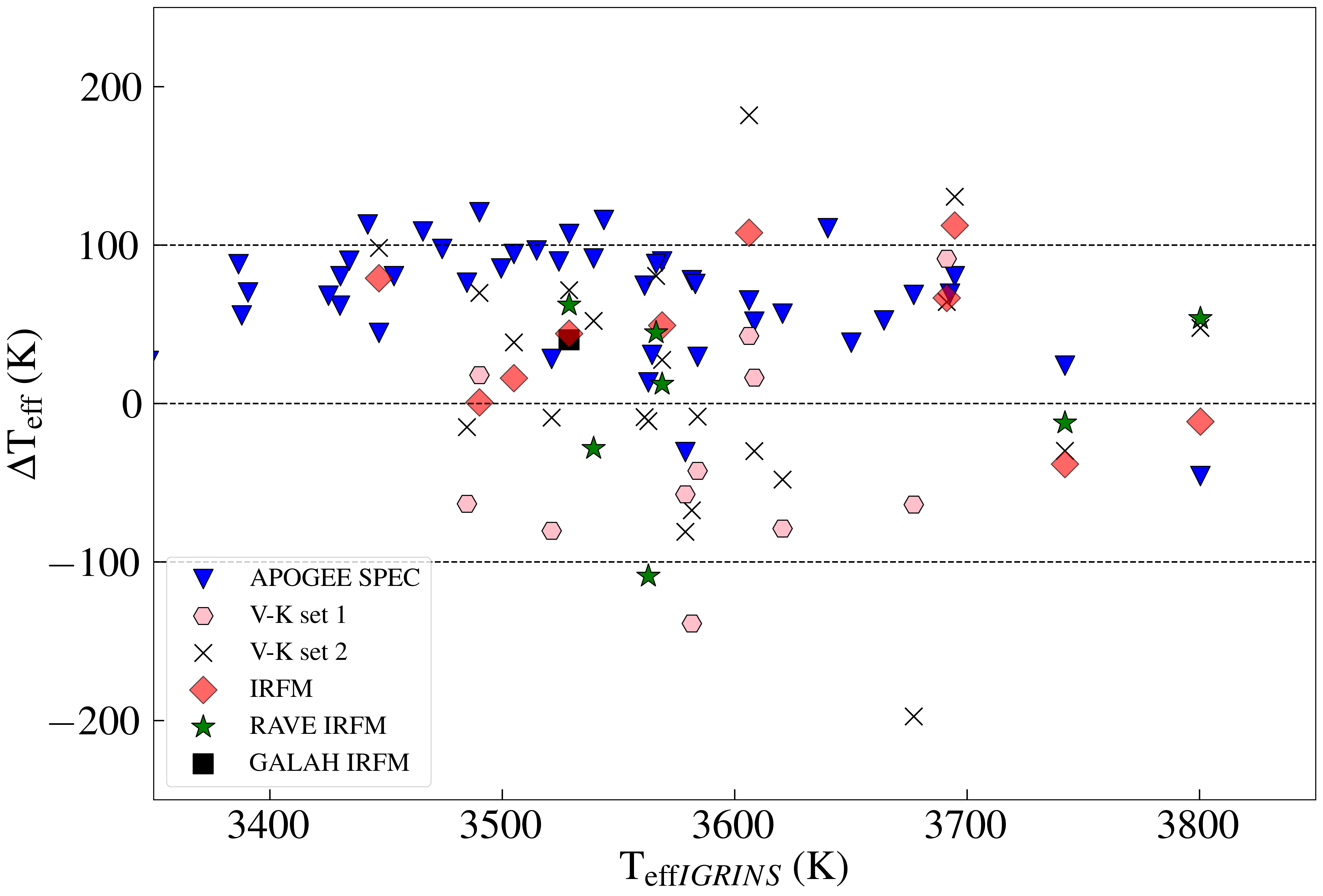}
  \caption{The differences (GALAH/RAVE/APOGEE/(V-K)/IRFM - our method) in \teff\, as a function of the \teff\, derived using our method on the x-axis. The black dashed lines indicate the difference values of -100 K, 0 K and +100 K. Different colored symbols represent values from different sources: blue inverted triangles - APOGEE spectroscopic estimates, black squared -  GALAH IRFM estimates, green star symbols - RAVE IRFM estimates, red diamonds - our IRFM \teff\, estimates, and pink hexagons - \teff\, estimated using \teff\, versus (V-K) relation from \cite{Bessell:1998} with extinctions in V and K based on the E(B-V) values from APOGEE.   }
  \label{fig:teffdiff}%
\end{figure*}

\subsubsection{Nearby M giants}
\label{sec:nearbyM}

As a test case of our method, we selected six nearby M giants, the H- and K-band spectra of which are available in the IGRINS spectral library \citep{Park:2018,rrisa}. We compiled their \teff, \logg, and \feh, derived using different methods from various literature sources. \teff\, for four of these stars (HD 132813, HD 89758, HD 175588, and HD 224935) were estimated based on the very precise angular diameter measurements using the Navy Precision Optical Interferometer (NPOI) in combination with distances from Gaia parallaxes \citep{Baines:2021}. Hence these values of \teff\, are very accurate and reliable. Further, we also have \teff\ for the four stars (HD 132813, HD 175588, HD 224935, and HD 96360) from \cite{Lebzelter:2019}, estimated using the \teff–(V–K) relation derived from a homogeneous set of angular diameters by \cite{Richichi:1999}. They also estimated \logg\, for one star (HD 132813) using the fundamental relation of \logg\, assuming a mass of 1.2 M$_{\odot}$, and luminosity and distances from Gaia. Unlike the above mentioned studies, \cite{Sharma:2016} provides all three stellar parameters for five stars (HD 132813, HD 89758, HD 175588, HD 101153, and HD 96360) based on full spectrum fitting of 331 stars in the MILES and ELODIE spectral libraries making use of the newer version, V2, of the MILES interpolator and the spectrum fitting tool, ULySS. For one star (HD 89758), we found all three stellar parameters in \cite{Jonsson:2014} with the \teff\, provided from angular diameter measurement by \cite{Mozurkewich:2003}, \logg\, estimated using the fundamental relation, and \feh\  determined from Fe I lines in a visual spectrum from the ELODIE spectral archive. For HD96360, \cite{Smith:1990} estimated \teff\ based on the \teff-spectral-type relation from \citet{Tsuji:1981}, \logg\ using the fundamental relation assuming a mass of 1.5 M$_{\odot}$, and \feh\, as the mean of the Ti, Fe and Ni abundances. These parameters were adopted by \cite{Guerco:2019}, who also provide all three stellar parameters for HD 101153 with, in this case, the \teff\ estimated using the \teff-(V-K)$_{0}$ relation from \cite{Bessell:1998}, \logg\, using the PARAM 1.3 code, and \feh\ determined using a sample of 19 Fe I lines in the K-band wavelength regime. Finally, for all six stars, we determined \teff\ using the following \teff-(V-K)$_{0}$ relation from \cite{Bessell:1998}:

\begin{equation}
\begin{aligned}
    T_\mathrm{eff} &= 9102.523 - 3030.654 (\mathrm{V} - \mathrm{K} )_{0} + 633.3732 (\mathrm{V} - \mathrm{K} )_{0}^{2} \\
    & - 60.73879 (\mathrm{V} - \mathrm{K} )_{0}^{3} + 2.135847(\mathrm{V} - \mathrm{K} )_{0}^{4}
\end{aligned}
\label{eqn:bessel}
\end{equation}

By using the python package dustmaps\footnote{https://dustmaps.readthedocs.io/en/latest/} we estimated the  dereddened values of V and K for this equation, with the reddening values (E(B-V)) from the two dimensional map of dust, constructed by \citep{SFD:1998} based on far-infrared emission of dust. We use the conversion factors of 3.07 and 0.366  to determine the extinctions A$_{V}$ and A$_{K}$ respectively (see Table A1 in \citet{Casagrande:2014}).

We also determined the stellar parameters for these stars with our method using IGRINS H-band spectra and assuming that the stars belong to the thin disc stellar population. Table~\ref{table:standard} lists the stellar parameters for all six stars determined with our method along with the V, K$_{2MASS}$ and E(B-V) values, as well as the stellar parameters from the literature. In the Figure~\ref{fig:benchmark}, we plot the difference in stellar parameters (literature - IGRINS) as a function of our IGRINS parameters. The differences in \teff\, for comparisons with multiple literature sources, lie well within $\pm$100 K, with a minimum difference seen when compared to \teff\, derived using the \cite{Bessell:1998} relation for all six stars. The differences in \logg\, are largest for HD96360 and HD89758 when compared to the values from \cite{Sharma:2016} with uncertainties $\sim$ 0.2 dex. We would like to point out that the rest of the stars with available \logg\, have small differences that lie within 0.1 dex. The differences in \feh\, also lie within levels of $\pm$0.2 dex which is also comparable to the uncertainties in the measurements in the literature. Thus overall, there is a very good agreement for all three stellar parameters we determined using our method for the six well studied nearby M giants. This clearly is an indication of the efficiency of our method to determined accurate stellar parameters for M giants.

\begin{table*}
\caption{ Stellar parameters, [C/Fe] and [N/Fe] values of each star determined in this work along with their assumed stellar population and [O/Fe] based on the APOGEE [Mg/Fe] versus [Fe/H] trend. }\label{table:all}
\begin{tabular}{c c c c c c c c c c}
\hline
\hline
 Index  & Population  & T$_\mathrm{eff}$ & $\log g$  & [Fe/H]  &  $\xi_\mathrm{micro}$   & [C/Fe] & [N/Fe] & [O/Fe] & Cross match  \\
 \hline
 & & K & log(cm/s$^{2}$) & dex & km/s & dex & dex & dex & \\
\hline

1 & thin  &  3490  &  0.48  &  -0.28  &  2.03  &  -0.02  &  0.17  &  0.1  &  -- \\  
2 &  thin  &  3694  &  0.74  &  -0.45  &  1.88  &  0.02  &  0.19  &  0.16  &  --  \\  
3 &  thick  &  3562  &  0.48  &  -0.51  &  2.14  &  0.21  &  0.3  &  0.38  &  RAVE  \\  
4 &  thick  &  3742  &  1.08  &  0.0  &  1.78  &  0.08  &  0.29  &  0.15  &  RAVE  \\  
5 & thin  &  3677  &  0.92  &  -0.07  &  1.78  &  -0.12  &  0.21  &  0.04  &  --  \\  
6 &   thin  &  3692  &  0.68  &  -0.54  &  2.01  &  0.02  &  0.31  &  0.19  &  --  \\  
7 &   thin  &  3583  &  0.42  &  -0.66  &  2.39  &  0.05  &  0.33  &  0.24  &  --  \\  
8 &   thick  &  3620  &  0.65  &  -0.38  &  2.1  &  0.17  &  0.22  &  0.33  &  --  \\  
9 &   thin  &  3608  &  0.55  &  -0.52  &  2.24  &  0.06  &  0.27  &  0.19  &  --  \\  
10 &  thick  &  3800  &  0.59  &  -0.92  &  2.72  &  0.21  &  0.37  &  0.56  &  RAVE  \\  
11 &   thin  &  3521  &  0.4  &  -0.52  &  2.19  &  0.0  &  0.31  &  0.18  &  --  \\  
12 &   thin  &  3484  &  0.32  &  -0.55  &  2.09  &  0.04  &  0.3  &  0.2  &  --  \\  
13 &   thin  &  3581  &  0.67  &  -0.21  &  2.15  &  -0.09  &  0.31  &  0.08  &  --  \\  
14 &   thin  &  3606  &  0.52  &  -0.56  &  1.96  &  0.04  &  0.27  &  0.2  &  --  \\  
15 &   thin  &  3561  &  0.56  &  -0.36  &  2.24  &  -0.08  &  0.41  &  0.13  &  --  \\  
16 &   thin  &  3568  &  0.96  &  0.25  &  1.83  &  -0.18  &  0.39  &  -0.08  &  RAVE, GALAH$^{*}$  \\ 
17 &   thin  &  3566  &  0.51  &  -0.46  &  2.02  &  0.01  &  0.24  &  0.16  &  RAVE  \\  
18 &   thin  &  3539  &  0.5  &  -0.41  &  2.05  &  0.03  &  0.23  &  0.15  &  RAVE, GALAH$^{*}$  \\  
19 &   thin  &  3528  &  0.61  &  -0.15  &  1.92  &  -0.02  &  0.13  &  0.06  &  RAVE, GALAH  \\  
20 &   thin  &  3504  &  0.61  &  -0.08  &  1.81  &  0.01  &  0.16  &  0.03  &  --  \\  
21 &   thin  &  3474  &  0.69  &  0.12  &  1.94  &  -0.07  &  0.21  &  -0.04  &  --  \\  
22 &   thin  &  3543  &  0.8  &  0.11  &  1.95  &  -0.05  &  0.2  &  -0.04  &  --  \\  
23 &   thin  &  3386  &  0.55  &  0.13  &  1.82  &  -0.04  &  0.14  &  -0.05  &  --  \\  
24 &   thin  &  3387  &  0.52  &  0.08  &  1.92  &  -0.03  &  0.21  &  -0.03  &  --  \\  
25 &   thin  &  3453  &  0.63  &  0.08  &  1.91  &  -0.1  &  0.22  &  -0.03  &  --  \\  
26 &   thin  &  3465  &  0.62  &  0.04  &  1.83  &  -0.09  &  0.33  &  -0.02  &  --  \\  
27 &   thin  &  3430  &  0.54  &  0.0  &  1.95  &  -0.02  &  0.16  &  -0.0  &  --  \\  
28 &   thin  &  3499  &  0.62  &  -0.06  &  2.01  &  -0.07  &  0.25  &  0.02  &  --  \\  
29 &   thin  &  3639  &  0.89  &  -0.0  &  1.76  &  -0.05  &  0.08  &  0.01  &  --  \\  
30 &   thin  &  3524  &  0.56  &  -0.25  &  1.98  &  0.02  &  0.15  &  0.09  &  --  \\  
31 &   thin  &  3664  &  1.11  &  0.23  &  1.99  &  -0.09  &  0.35  &  -0.08  &  --  \\  
32 &   thin  &  3430  &  0.55  &  0.02  &  1.92  &  -0.09  &  0.21  &  -0.01  &  --  \\  
33 &   thin  &  3425  &  0.54  &  0.02  &  1.87  &  -0.07  &  0.25  &  -0.01  &  --  \\  
34 &   thin  &  3442  &  0.68  &  0.18  &  1.85  &  -0.0  &  0.16  &  -0.06  &  --  \\  
35 &   thin  &  3514  &  0.53  &  -0.26  &  2.02  &  -0.04  &  0.16  &  0.09  &  --  \\  
36 &   thin  &  3446  &  0.61  &  0.08  &  1.99  &  -0.04  &  0.21  &  -0.03  &  --  \\  
37 &   thin  &  3650  &  0.98  &  0.1  &  1.8  &  -0.11  &  0.29  &  -0.04  &  --  \\  
38 &   thin  &  3582  &  0.96  &  0.23  &  1.8  &  -0.11  &  0.31  &  -0.09  &  --  \\  
39 &   thick  &  3691  &  0.76  &  -0.4  &  1.98  &  0.17  &  0.14  &  0.34  &  --  \\  
40 &   thin  &  3564  &  0.95  &  0.25  &  2.2  &  -0.04  &  0.38  &  -0.09  &  --  \\  
41 &   thin  &  3347  &  0.46  &  0.09  &  1.98  &  -0.09  &  0.21  &  -0.03  &  --  \\  
42 &   thin  &  3390  &  0.48  &  0.01  &  1.96  &  -0.01  &  0.27  &  -0.0  &  --  \\  
43 &   thin  &  3434  &  0.59  &  0.07  &  1.93  &  -0.07  &  0.24  &  -0.02  &  --  \\  
44 &   thick  &  3578  &  0.45  &  -0.59  &  2.26  &  0.08  &  0.54  &  0.41  &  GALAH$^{*}$  \\ 

\hline
\hline
\end{tabular}
\tablefoot{\\ *: No IRFM \teff\, in GALAH DR3 }
\end{table*}

\begin{figure*}
  \includegraphics[width=\textwidth]{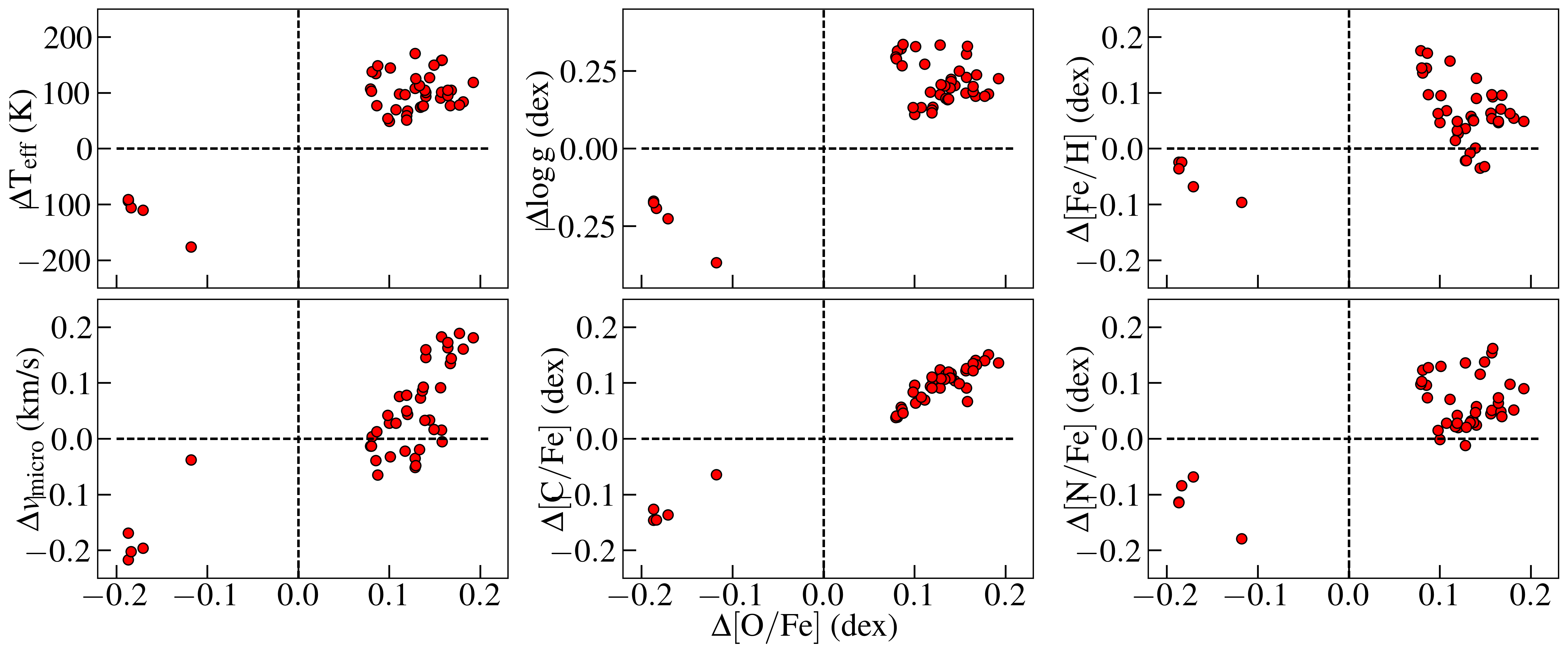}
  \caption{The difference in \teff, \logg, \feh, $\xi_\mathrm{micro}$, [C/Fe], and [N/Fe] as a function of the difference in [O/Fe] resulting from the change of population assumption from thin to thick disc and vice versa for the 44 solar neighborhood M giants. Circle markers with positive $\delta$[O/Fe] values represent actual thin disc stars that are assumed to be thick disc for this exercise.   }
  \label{fig:uncertainty}%
\end{figure*}

\begin{figure}
  \includegraphics[width=\columnwidth]{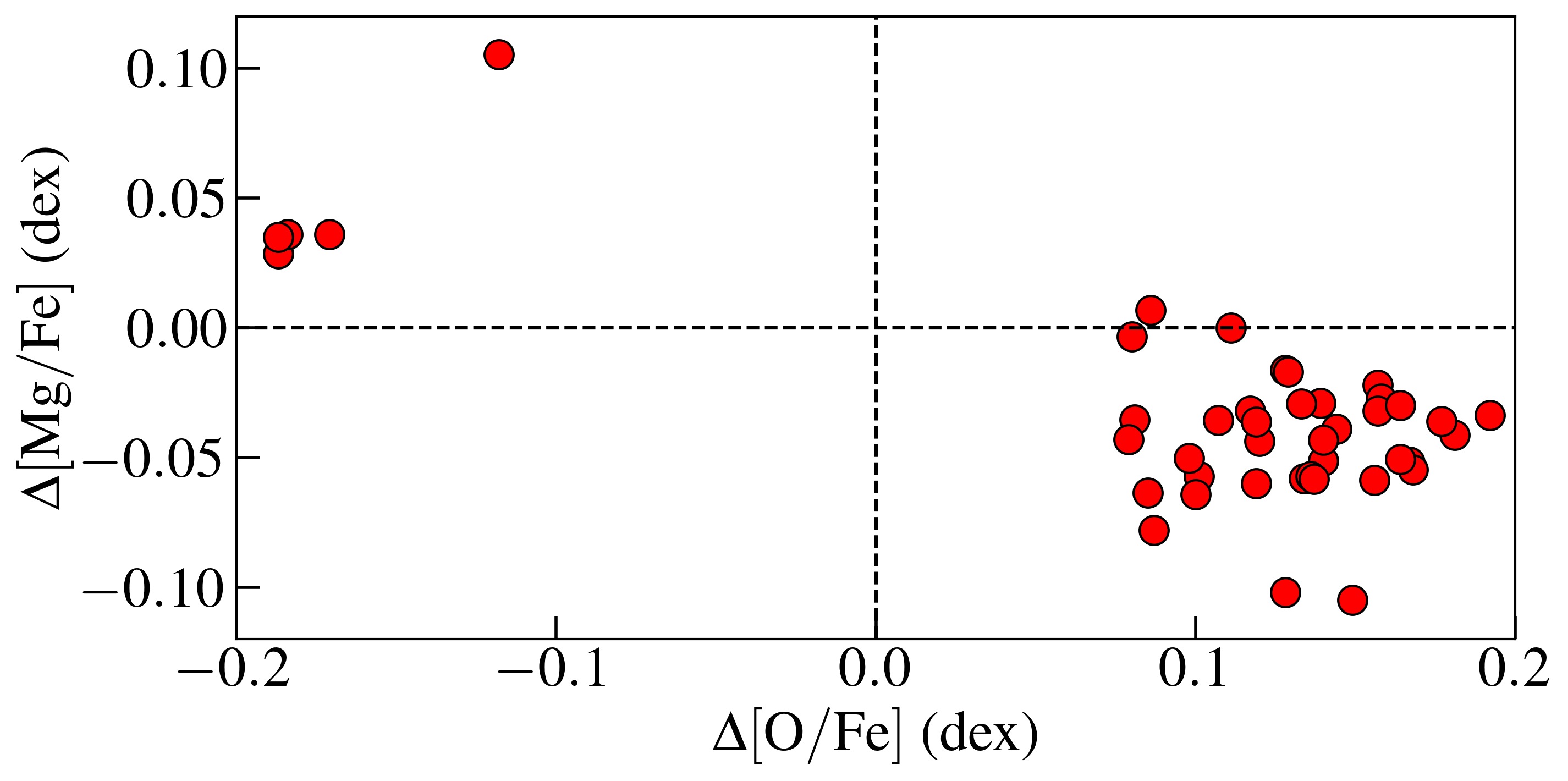}
  \caption{Same as Figure~\ref{fig:uncertainty} except for the difference in [Mg/Fe] plotted on y-axis. It is evident that [Mg/Fe] decreases for thin disc stars wrongly assumed to be thick disc and [Mg/Fe] increases for thick disc stars wrongly assumed to be thin disc. This demonstrates that a wrong classification would be caught in the derived alpha abundances, and a corrective iteration of our method can be done.}
  \label{fig:mgdiff}%
\end{figure}

\subsubsection{Solar neighbourhood M giants}
\label{sec:SNM}

Next, we applied our method to our new IGRINS spectra of the 44 solar neighbourhood M giants. All 44 stars have also been observed by the APOGEE survey with their stellar parameters and individual elemental abundances available in the latest data release, DR17, for all stars except one ($\#$39). We identified five stars belonging to the thick disc population based on their enhanced APOGEE magnesium abundance from the [Mg/Fe] versus [Fe/H] plot. Thus, we fixed the oxygen abundance for these five stars according to the thick-disc trend and for the remaining 40 stars according to the thin-disc trend in  Figure~\ref{fig:Otrend}. Later, based on our abundance analysis (alpha abundances), $\#$39 was identified to be a thick-disc star. When we re-estimated the parameters assuming the thick-disc oxygen trend (increase of 0.17 dex in [O/Fe]), we found a 110 K increase in \teff\,,  0.23 dex increase in \logg\,, 0.07 dex decrease in \feh\,, 0.2 km/s increase in $\xi_\mathrm{micro}$, 0.14 dex increase in [C/Fe], and a 0.07 dex increase in [N/Fe]. It is surprising, and reassuring,  that even after the initial misidentification, we were able to rightly deduce the correct population based on the abundances we derived using the wrong stellar parameters. Nevertheless, it is an encouraging sign that there are ways to identify and correct such wrong initial assumptions, see further discussion in the section on uncertainties, Section \ref{sec:uncertain}.

APOGEE derived \teff, \logg, \feh, $\xi_\mathrm{micro}$, $\xi_\mathrm{macro}$, [C/M], [N/M], [O/M], and general alpha abundance, [$\alpha$/M], simultaneously using the APOGEE Stellar Parameter and Chemical Abundance Pipeline (ASPCAP; \citet{ASPCAP}) by interpolating in a pre-computed grid of synthetic spectra and finding the best-fitting stellar parameters that describe an observed spectrum. In addition to the raw, spectroscopic parameters and abundances, APOGEE also provides calibrated values.  
%{\color{red}{Govind: yes, maybe get Henrik say saomething? Should we talk more about the APOGEE calibrations and why we are not using them}}. 
Since the uncalibrated spectroscopic parameters and abundances are directly derived from the APOGEE spectra, in this work, we make use of spectroscopic parameters (reported under the ASPCAP output array FPARAM) to compare with the values we derived using our method.

In figure~\ref{fig:paramdiff}, we plot the difference in the parameters and abundances (our method - APOGEE) as a function of APOGEE uncalibrated values. We estimated the mean value of the differences and scatter (mid value of 84$^{th}$ and 16$^{th}$ percentiles) which are listed in each panel. The \teff, estimated using our method is lower than APOGEE's \teff, for all stars except two: $\#$10 and $\#$44. In general, the difference in \teff\, is small with a mean of -67 K and scatter of 33 K, thus lying within 100 K for the majority of the stars with no significant trends. Similar to \teff, the surface gravity, \logg, from our method is lower compared to APOGEE with a mean difference of -0.31 dex and scatter of 0.15 dex. We also see a trend of lower differences ($<$ -0.1 dex) at lower values of APOGEE \logg\, that increases to $\sim$ -0.5 dex with larger scatter at higher values of APOGEE \logg. It has been pointed out in \cite{Holtzman:2018} that the  spectroscopic (raw) \logg\, values for giants determined by APOGEE ASPCAP are systematically higher than those derived from asteroseismology. This could possibly explain the consistently lower difference we find in \logg. At the same time, currently available measurements of asteroseismic \logg\, are limited to stars with \teff\, $>$ 3800 K \citep{Pinsonneault:2018}. Our metallicities are in agreement with APOGEE metallicities with a mean difference of 0.02 dex and a small scatter of 0.05 dex. There is a hint of change in the trend in the metallicity differences, from positive to negative, at higher metallicities, but this cannot be confirmed with the current sample. The difference in $\xi_\mathrm{micro}$ shows a clear trend, with higher values using our method for lower APOGEE $\xi_\mathrm{micro}$ but tends to agree at higher APOGEE $\xi_\mathrm{micro}$. Our $\xi_\mathrm{micro}$ for a majority of the stars lie in a narrow range of 1.8-2.4 km/s and a high value of 2.7 km/s for the most metal poor star, $\#$10. It is encouraging that our values of $\xi_\mathrm{micro}$ are  in reasonable range of values usually accepted for giants \citep{Smith:2013}. Finally, on average we find small differences for [C/Fe], [N/Fe] and [O/Fe] with respect to APOGEE. We note that [C/Fe] from our method are higher for the five thick disc stars similar to what APOGEE finds.  

On further cross match carried out with other large spectroscopic surveys, we found seven stars in the RAVE survey catalog (DR6; \citet{Steinmetz:2020}) and four stars in the GALAH survey catalog (DR3; \citet{Buder:2021}). Based on the quality flags, algo$\_$conv$\_$madera (set to 3) in RAVE and flag$\_$sp (not set to 0) in GALAH, these stars do not have reliable stellar parameters determined from their spectra in both the survey catalogs. Meanwhile, the photometric \teff\, has been estimated using the IRFM method \citep{Casagrande:2010} for all five stars in RAVE and for one star in GALAH \citep{Casagrande:2021}. 

In addition to comparing against APOGEE's values, we estimated photometric effective temperatures using the \teff-(V-K) relation (Equation~\ref{eqn:bessel}). In doing so, we explored the effect of different reddening values. For 11 stars we had E(B-V) from APOGEE directly (set 1, hexagons in Fig \ref{fig:teffdiff}), whereas for 21 stars (set 2, crosses in Fig \ref{fig:teffdiff}) we adopted reddening values from \citet{SFD:1998} renormalized as described in \cite{Casagrande:19}. The input photometry being the same, this comparison is a sobering example of how reddening alone can easily introduce a scatter of order $\pm50$~K for our sample.  

In addition to using the above colour relation, we also implemented Gaia DR3 and 2MASS photometry in the IRFM as per \cite{Casagrande:2021} adopting our spectroscopic values of $\log g$ and [Fe/H] and our renormalized reddening values. To better gauge into the uncertainties, for each star we MonteCarlo the errors into the IRFM, adopting quoted uncertainties for input stellar parameters and photometry, and allowing for 10 percent uncertainty in reddening. For stars with E(B-V)<0.85, our MonteCarlo uncertainties are within 100~K, but they linearly increase to several hundreds of K for higher values of reddening. In order to retain stars with reliably determined IRFM temperatures we thus restrict ourselves to stars with E(B-V) below the above threshold, which correspond to \teff uncertainties in the range 30 to 60 K. From Figure \ref{fig:teffdiff} it is clear that the IRFM values (red diamonds) are consistent with our spectroscopic determinations, the mean difference being 42 K with a standard deviation of 48 K. We could not estimate IRFM temperatures for the six nearby M giants owing to their extreme brightness.

Figure \ref{fig:teffdiff} also shows the comparison with the IRFM effective temperatures from RAVE DR6 \citep{Steinmetz:2020}. In this version of the IRFM APASS photometry was used instead, along with $\log g$ and [Fe/H] derived from RAVE. This comparison shows the effect of the adopted stellar parameters and photometry, with the caveat that Gaia DR3 is far superior than APASS. Finally we also show the comparison with one stars with IRFM effective temperature from GALAH which was based instead on the Gaia DR2 photometry and transmission curves.

From the different values and methods compared in Figure~\ref{fig:teffdiff} we can thus conclude that our effective temperature determinations are reliable especially in comparison with several other ones available in the literature. In particular, the scatter with respect to several photometric determinations points to the uncertainty introduced by the limited precision to which reddening can be estimated. 
Since our determined effective temperatures agree excellently with the six nearby, well studied M giants in the IGRINS spectral library (in Fig. \ref{fig:benchmark}), we have good reasons to believe that our values are more accurate. Especially, the moderately tight offset againtst \teff\ from APOGEE points to good internal precision once a systematic offset of about 100 K for stars cooler than 3700 K is removed.   

% We found only one star out of the 8 stars with photometric \teff\, from (V-K) relation showing difference outside the -100 K difference. 

\begin{figure*}
  \includegraphics[width=\textwidth]{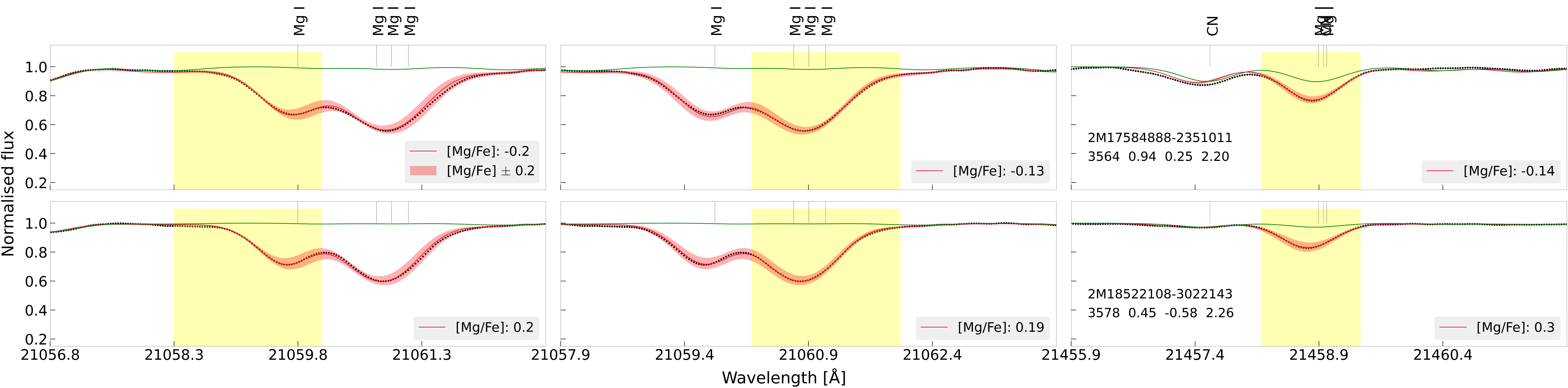}
  \caption{Wavelength regions centered at the five selected Magnesium lines for the thin disc star 2M17584888-2351011 ($\#$40; top row panels) and the thick disc star 2M18522108-3022143 ($\#$44; bottom row panels) with the panels in each row arranged in the increasing order of wavelengths of selected lines. In each panel, the black circles denote the observed spectrum, crimson line denote the best fit synthetic spectrum and the red band denote the variation in the synthetic spectrum for $\pm$ 0.2 dex difference in the [Mg/Fe]. The yellow bands in each panel represent the line masks defined for the Mg lines wherein SME fits observed spectra by varying Magnesium abundance and finds the best synthetic spectra fit by chi-square minimisation. The green line shows the synthetic spectrum without Mg, also indicating any possible blends in the line. The [Mg/Fe] values corresponding to the best fit case for each Mg line is listed in each panel. All identified atomic and molecular lines are also denoted in the top part of the top row panels.}    
  \label{fig:mgspectra}%
\end{figure*}

\begin{figure*}
  \includegraphics[width=\textwidth]{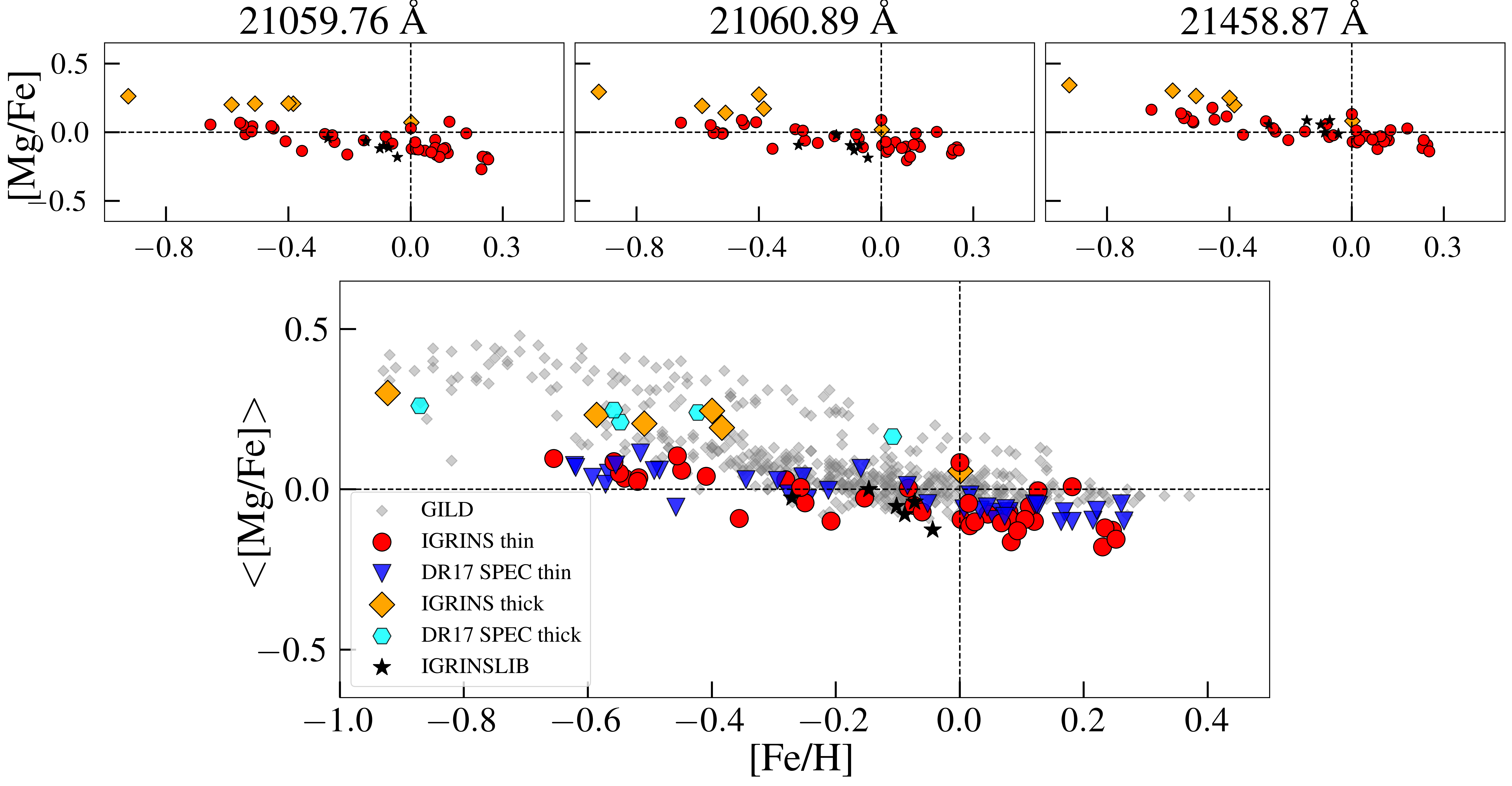}
  \caption{The [Mg/Fe] versus [Fe/H] trends estimated from each Magnesium line (top row panels) and the trend from the mean of line-by-line [Mg/Fe] estimates (bottom panel). The red circles (thin disc), orange diamonds (thick disc) and black star symbols (thin disc) in all panels represent the 44 solar neighborhood M giants and the six nearby M giants respectively. Gray diamonds in the bottom panel represent the stars in the Giants in the Local Disc (GILD) sample with parameters and abundances estimated from optical FIES spectra and shifted down by 0.1 dex. Blue inverted triangles (thin disc) and cyan hexagons (thick disc) represent the APOGEE spectroscopic estimates for the 43 solar neighborhood stars. }
  \label{fig:mgindiv}%
\end{figure*}

\subsection{Uncertainties in the derived stellar parameters}
\label{sec:uncertain}

As explained in the section~\ref{sec:parameters}, our method to estimate stellar parameters depends largely on the oxygen abundance which we fix based on the metallicity of the star and the Milky Way population to which the star belongs (see Figure~\ref{fig:Otrend}). Thus, the assumption of the population or the oxygen abundance may be considered as the main source of uncertainty for our stellar parameters. Hence, in order to estimate typical uncertainties in the parameters and abundances we derive, we implemented our method assuming the 38 thin disc stars in the solar neighbourhood sample as thick disc stars and vice versa for the 6 thick disc stars. This is similar to the mis-classification of the star $\#$39 as thin disc star in the section~\ref{sec:SNM}. 

The results of this exercise are shown in the Figure~\ref{fig:uncertainty} with the differences in oxygen abundance on the x-axis and the differences in six other stellar parameters on the y-axis in each panel. In this figure, actual thick disc stars have negative $\Delta$[O/Fe], since [O/Fe] is lower for the thin disc trend and the actual thin disc stars have positive $\Delta$[O/Fe]. Assuming lower oxygen abundance for the most metal poor star, $\#$10, resulted in stellar parameters that were outside our adopted grid limits. Hence, we omit this star in this exercise. For the maximum difference of $\pm$0.2 dex in [O/Fe], we found typical differences of $\pm$200 K in \teff\,, $\pm$0.25 dex in \logg\,, $\pm$0.2 km/s in $\xi_\mathrm{micro}$, $\pm$0.2 dex in [C/Fe], and $\pm$0.2 dex in [N/Fe]. The difference in metallicity is found to be closer to 0 for [O/Fe] differences of +0.2 and -0.2 dex, but larger (up to $\sim$+0.2 dex) in the range 0.0 $<$ $\Delta$[O/Fe] $<$ 0.2 dex. Thus, assuming a typical uncertainty of 0.15 dex in [O/Fe], the parameters estimated using our method will have typical uncertainties of $\pm$100 K in \teff\,, $\pm$0.2 dex in \logg\,, $\pm$0.1 dex in \feh\,, $\pm$0.1 km/s in $\xi_\mathrm{micro}$, $\pm$0.1 dex in [C/Fe], and $\pm$0.1 dex in [N/Fe]. But again, as we demonstrated for the star $\#$39  in the section~\ref{sec:SNM}, a wrong classification would be caught in the derived alpha abundances, and a corrective iteration can be done. To make sure this was not a chance occurrence, we determined [Mg/Fe] for all stars using stellar parameters determined based on a mis-classification. We found that the [Mg/Fe] abundances decreases further for the wrongly classified thin disc star and increases further for the wrongly classified thick disc star. This is demonstrated in the Figure~\ref{fig:mgdiff}. We will, therefore, safely be able to catch any miss-classification of a high- or low-alpha star by inspecting the derived [Mg/Fe] ratio.

%{\bf Shall we say anything on the errors if the stars would be younger than 3 Gyr? How large would the isochrone method give?}
Another source of uncertainty is the use of 10 Gyr YY isochrone to constrain \logg. As discussed in \cite{rich:17}, different age tracks for giants overlap which in turn should result in negligible difference in \logg\, if we choose a lower age isochrone track. Indeed, we find that \logg\, increases by only $\sim$ 0.1 dex if we use a 2 Gyr YY isochrone instead of the 10 Gyr one. 

Thus, we estimate the uncertainties in the derived stellar parameters using our method to be $\pm$100 K in \teff, $\pm$0.2 dex in \logg, $\pm$0.1 dex in \feh, and $\pm$0.1 km/s in $\xi_\mathrm{micro}$. Within these uncertainties, our parameters are in line with the comparison samples and methods presented in Figures \ref{fig:benchmark}-\ref{fig:teffdiff}.

\section{$\alpha$ Abundance Trends}

\label{sec:alphaabund}

Based on the stellar parameters determined using our iterative method, we determined elemental abundances of the following $\alpha$ elements: Mg, Si, Ca and Ti, for the six nearby M giants and 44 solar neighbourhood M giants. We adopt the solar abundance values for Mg (A(Mg)$_{\odot}$ = 7.53), Si (A(Si)$_{\odot}$ = 7.51), Ca (A(Ca)$_{\odot}$ = 6.31) and Ti (A(Ti)$_{\odot}$ = 4.90) from \cite{solar:sme}. In the following subsections, we discuss the individual and mean elemental abundance trends determined from a selected set of carefully selected absorption lines of each element in the H and K bands. For each element, we fitted the selected lines individually and determined the mean abundance value after removing lines that were too noisy, affected by spurious features, or affected by telluric lines that were not eliminated well enough in the telluric line removal procedure. %Finally, we chose the abundances from only those lines that are least affected by telluric lines as well as have good synthetic spectrum fits. 
From these chosen sets of abundances, we determined the mean abundance and a line-by-line scatter of each element for every star, see Tables~\ref{table:mgfe} to ~\ref{table:tife}. We will also compare the mean elemental abundance trends to the optical solar neighbourhood trends from {\it Giants In the Local Disk (GILD)} sample and to the APOGEE DR17 spectroscopic values if available. This comparison also serves as a way to further validate our stellar parameters and hence our method used to determine them.

In addition to the line-by-line scatter reported in the Tables~\ref{table:mgfe} to \ref{table:tife}, we determined the uncertainties in the elemental abundance estimates from each line that arise from the uncertainties in stellar parameters. As mentioned in Section~\ref{sec:uncertain}, parameters estimated using our method will have typical uncertainties of $\pm$100 K in \teff\,, $\pm$0.2 dex in \logg\,, $\pm$0.1 dex in \feh\,, $\pm$0.1 km/s in $\xi_\mathrm{micro}$. We selected seven stars with metallicities of $\sim$ -0.9 dex ($\#$10), -0.5 dex ($\#$11 and $\#$44), -0.25 dex ($\#$30), 0.0 dex ($\#$29), 0.1 dex ($\#$41) and 0.25 dex ($\#$40) to determine uncertainties. Thus we cover the entire metallicity range explored in this study. We randomly generated 50 sets of stellar parameters following a normal distribution with the stellar parameter value as the mean and these typical uncertainties as the standard deviation, and reanalyse each stellar spectrum using those parameters. We generated 50 sets of parameters for each of the seven stars. The resulting distribution of estimated abundances from each line is fitted with a Gaussian function. The dispersion estimated from this fit gives the uncertainty in abundances. Table~\ref{table:uncert_params} lists the uncertainties from the individual elemental lines as well as the mean uncertainty corresponding to each elemental abundance for these seven stars. The mean abundance uncertainties range from 0.04-0.08 dex for [Mg/Fe], 0.07-0.11 dex for [Si/Fe], 0.04-0.07 dex for [Ca/Fe] and 0.06-0.11 dex for [Ti/Fe].

\subsection{Magnesium (Mg)}
\label{sec:mg}

\begin{figure*}
  \includegraphics[width=\textwidth]{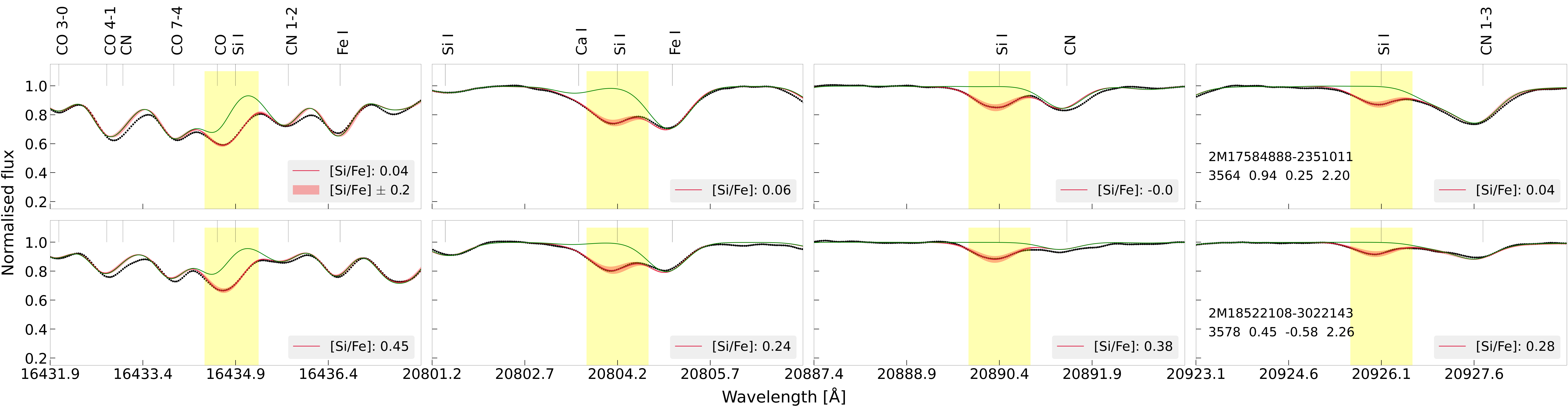}
  \caption{Similar to Figure~\ref{fig:mgspectra} but for silicon lines.}
  \label{fig:sispectra}%
\end{figure*}
 
\begin{figure*}
  \includegraphics[width=\textwidth]{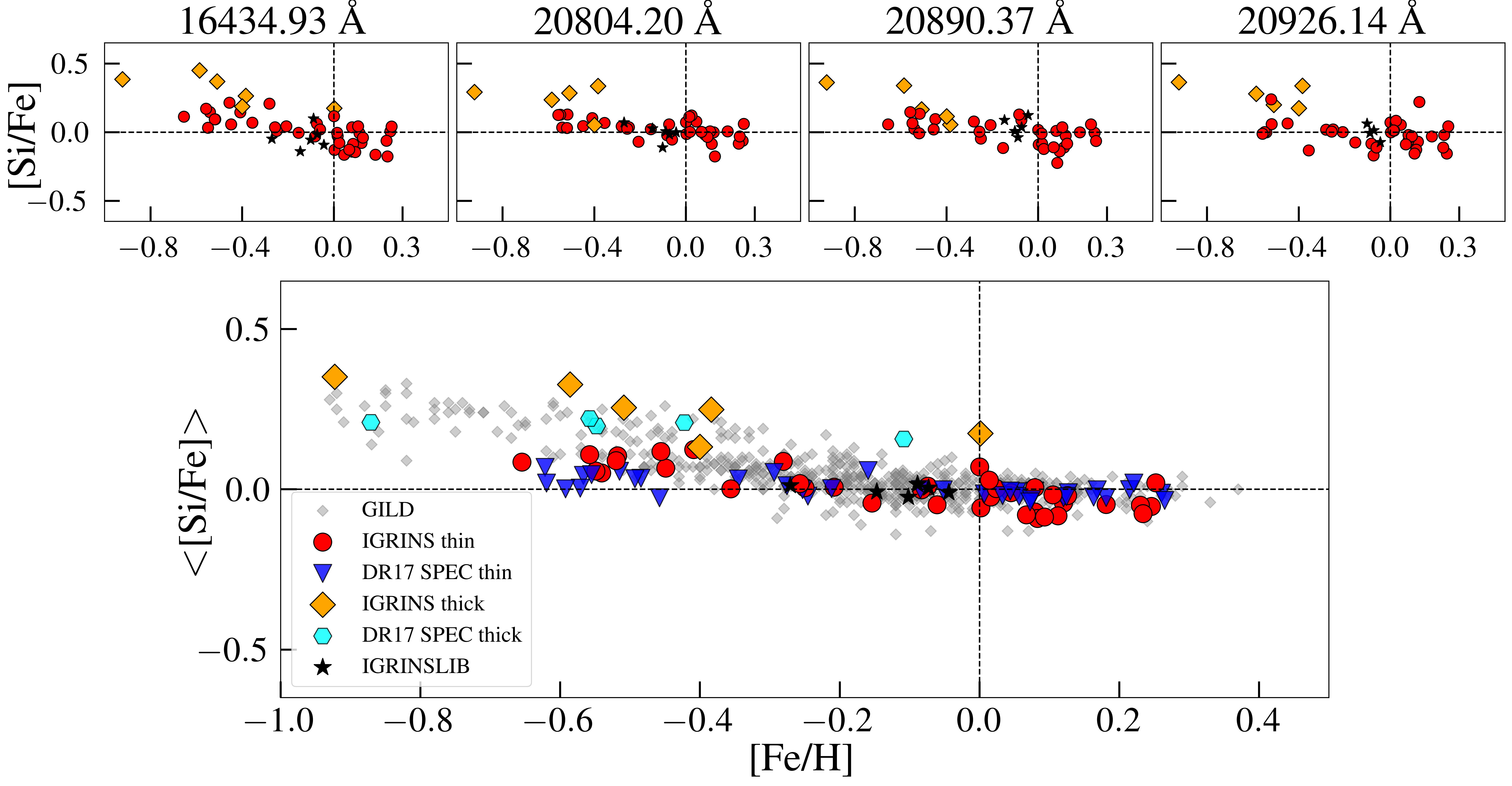}
  \caption{Similar to Figure~\ref{fig:mgindiv} but for [Si/Fe].}
  \label{fig:siindiv}%
\end{figure*}

\begin{figure*}
  \includegraphics[width=\textwidth]{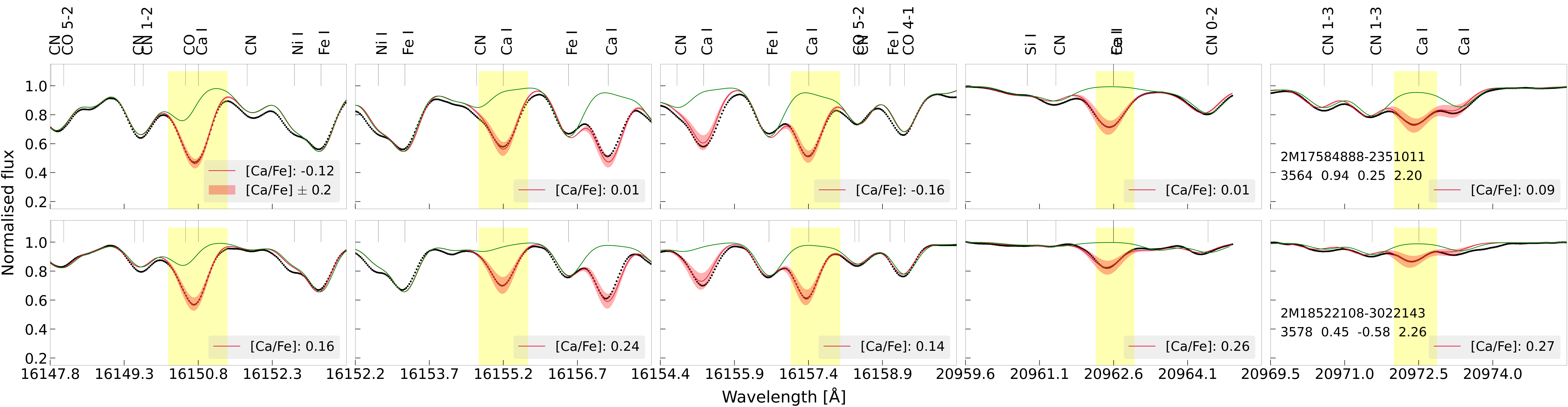}
  \caption{Similar to Figure~\ref{fig:mgspectra} but for calcium lines.}
  \label{fig:caspectra}%
\end{figure*}

\begin{figure*}
  \includegraphics[width=\textwidth]{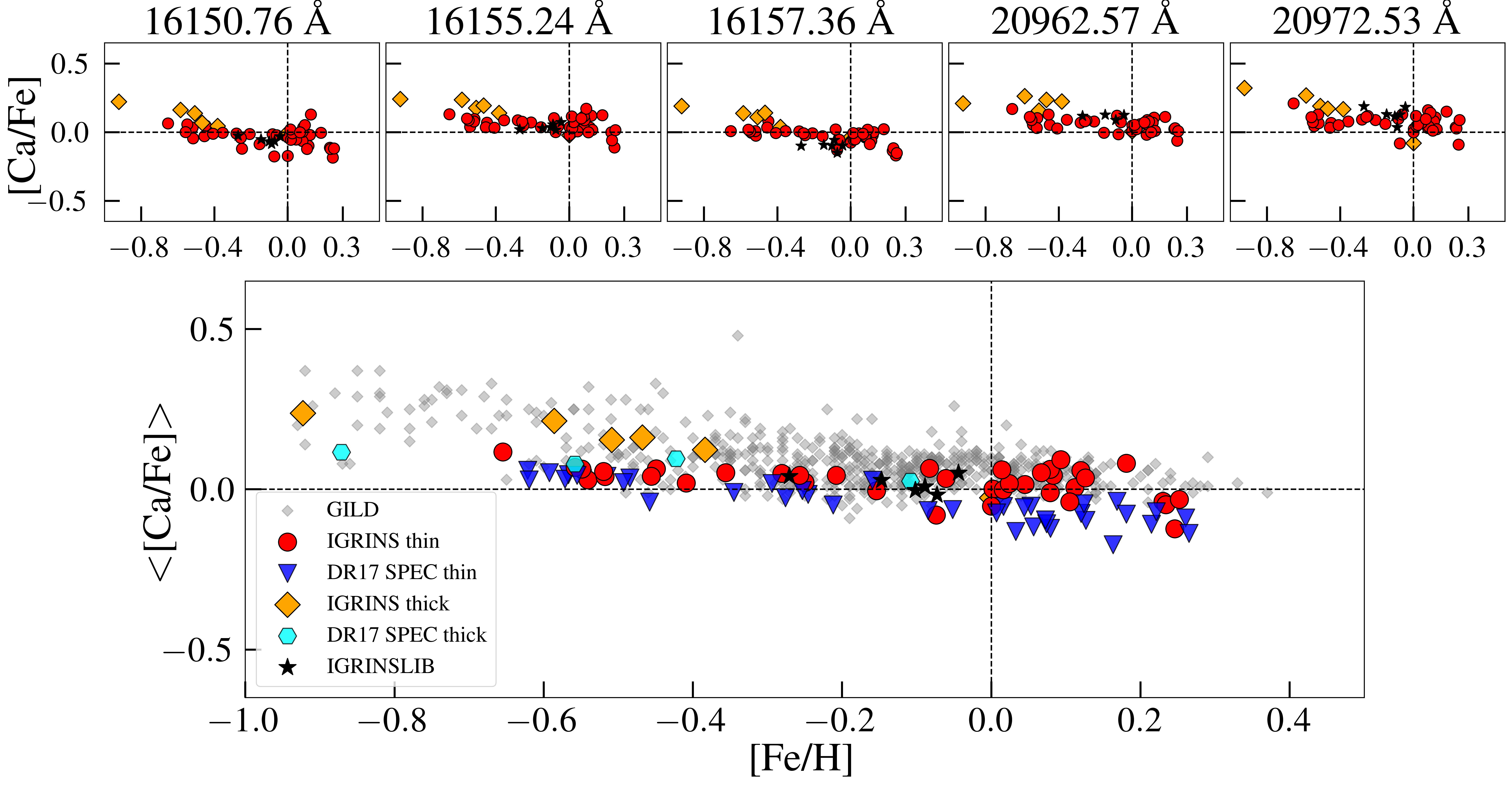}
  \caption{Similar to Figure~\ref{fig:mgindiv} but for [Ca/Fe]. The GILD [Ca/Fe] have been shifted down by 0.05 dex to normalize the comparison sample to the solar value.}
  \label{fig:caindiv}%
\end{figure*}

\begin{figure*}
  \includegraphics[width=\textwidth]{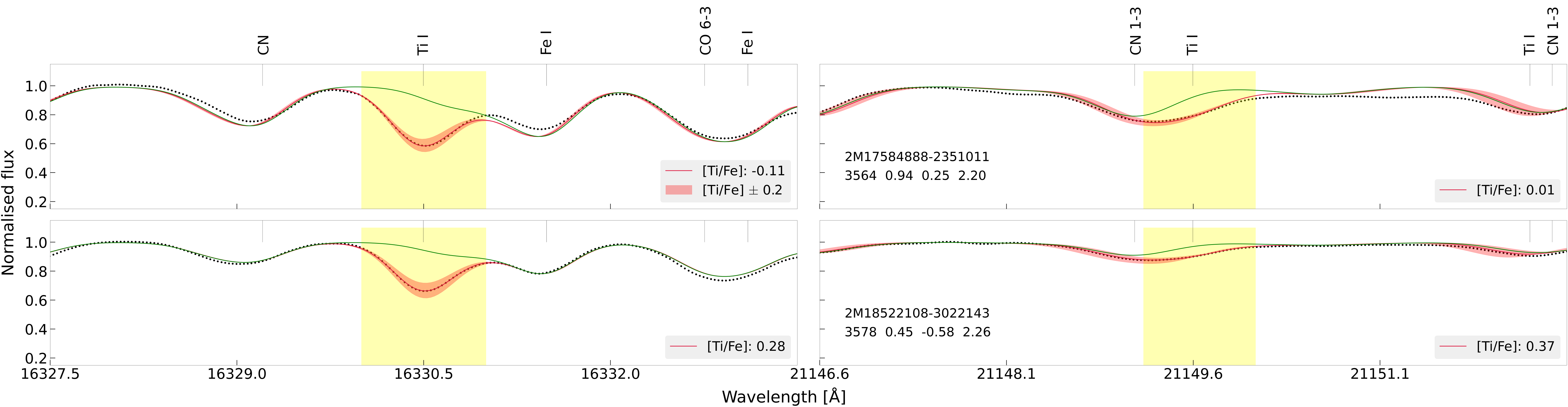}
  \caption{Similar to Figure~\ref{fig:mgspectra} but for titanium lines.}
  \label{fig:tispectra}%
\end{figure*}

\begin{figure*}
  \includegraphics[width=\textwidth]{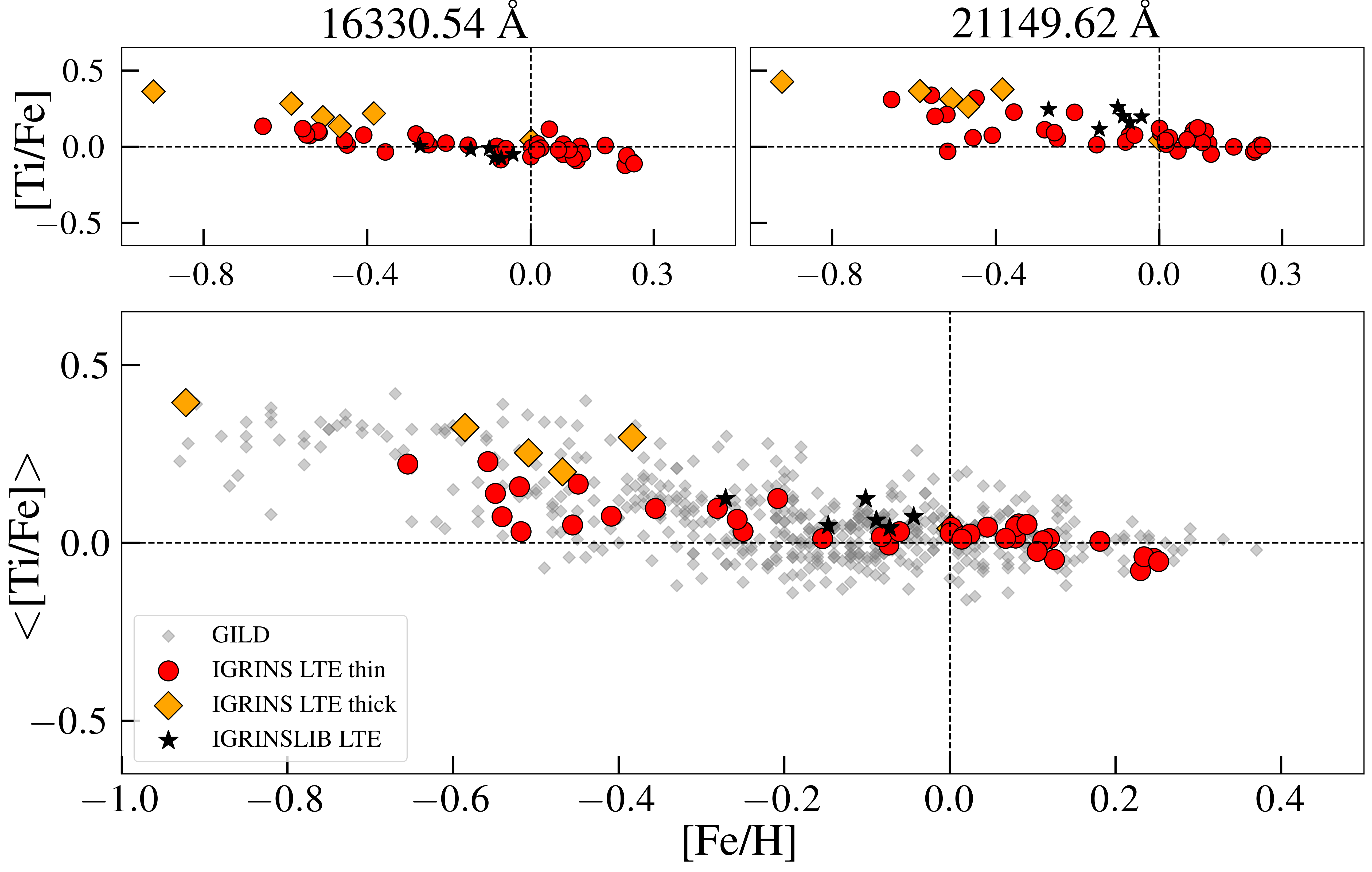}
  \caption{Similar to Figure~\ref{fig:mgindiv} but for [Ti/Fe].}
  \label{fig:tiindiv}%
\end{figure*}

% Among the five absorption lines from which we determined the magnesium abundances, two are in the H band: 15740.70 \AA, 15748.89 \AA, and three in the K band: 21059.76 \AA, 21060.89 \AA, 21458.87 \AA.

% The Mg abundances determined from the two H band lines at 15740.70 \AA\, and 15748.89 \AA\, were found to have strong correlation with the microturbulence (0.1-0.2 dex variation for $\Delta$$\xi_\mathrm{micro}$ = $\pm$ 0.2 dex) while insensitive to change in Mg abundances and thus are not used in this work.

We determined the magnesium abundances from three lines in the K band: 21059.76 \AA, 21060.89 \AA, 21458.87 \AA. The two H band Mg I lines at 15740.70 \AA\, and 15748.89 \AA\, could not be used in this work since we find these two lines to be saturated based on their insensitivity to $\pm$ 0.2 dex variation in Mg abundances. In addition, the Mg abundances from these two lines are found to have strong correlation with the microturbulence (0.1-0.2 dex variation for $\Delta$$\xi_\mathrm{micro}$ = $\pm$ 0.2 dex). 

In the Figure~\ref{fig:mgspectra}, we plot the selected Mg lines in the observed spectra of one thin disc star ($\#$40, top row panels) and one thick disc star ($\#$44, bottom row panels) in black circles, the synthetic spectrum fit to these lines in red, and the variation in the fit resulting $\pm$0.2 dex change in the abundance value as red band. The yellow bands in each panel represent the line masks defined for the Mg lines (avoiding the neighboring lines) wherein the SME fits observed spectra by varying Magnesium abundance and finds the best synthetic spectra fit by chi-square minimisation. The green line shows the synthetic spectrum without Mg, also indicating any possible blends in the line. The stellar parameters estimated for each star is listed in the left most panels along with the derived abundance from each line in the corresponding panel.

We plot the Mg abundance trend ([Mg/Fe] versus [Fe/H]) from each of these lines in the top five panels and the mean Mg abundance trend in the bottom panel of the Figure~\ref{fig:mgindiv}. In these plots, the red circles (thin disc) and orange diamonds (thick disc) represent the solar neighborhood sample and black star symbols represent the six nearby M giants. In the bottom panel, the GILD Mg abundance trend is plotted as gray diamonds while the APOGEE spectroscopic Mg values for the 43 solar neighborhood stars (except $\#$39) are plotted as blue inverted triangles (thin disc) and cyan hexagons (thick disc). 

% We found consistently higher ([Mg/Fe] $>$ 0.0) values for the Mg abundances determined from the two H band lines compared to the values derived from the three K band lines. As shown in the Figure~\ref{fig:mgspectra}, the two H band lines are stronger lines, which, being close to saturation, will make them very sensitive to the $\xi_\mathrm{micro}$ and not so much to abundance \citep{Gray:2008}.

 This is also evident in the negligible variation in the synthetic spectrum fit to the two H band lines for the $\pm$ 0.2 dex [Mg/Fe] variation in Figure~\ref{fig:mgspectra}. The [Mg/Fe] trend from all three K band lines show a decreasing trend with increasing metallcity, especially at supersolar metallicities \citep[as expected from the chemical evolution models, see ][]{matteucci:21} and a clear enhancement in [Mg/Fe] for thick disc stars compared to thin disc stars. [Mg/Fe] values from the multiplet lines 21059.76 \AA\, and 21060.89 \AA\, are found to have sub solar [Mg/Fe] at solar metallicity unlike those from the 21458.87 \AA\, that pass through the solar value as expected. This line also has a comparatively accurate $gf$-value as listed in NIST database with an accuracy grade of B+ (which means $\le 7\%$). The line parameters we used for the two multiplet K band lines have been found to result in good synthetic spectrum fit to the lines in the high resolution Sun and Arcturus spectra. \cite{Nieuwmunster:2023} using the same lines and line data, found similar low trend for inner bulge stars and warmer solar neighborhood K giants. We would also like to point out that the [Mg/Fe] values determined from each line for the six near by M giants are consistent with the corresponding trend obtained for the solar neighborhood stars.
%Thus further detailed investigation is needed to understand why the [Mg/Fe] trends from these lines give lower values compared to the expected trend.
% , but we found around $\sim$ 0.05 - 0.1 dex enhancement in our mean abundance for thin disc stars that also pass through the solar value at solar metallicity

We determined the mean [Mg/Fe] for each star from the lines which are deemed to have good synthetic spectrum fit by visual check and not affected by noise or telluric lines. The individual and mean [Mg/Fe] for each star along with the standard deviation value is listed in the Table~\ref{table:mgfe}. Our mean [Mg/Fe] trend shows a clear thin disc-thick disc dichotomy with enhanced values for thick disc stars. On comparison with the APOGEE DR17 spectroscopic [Mg/Fe] values for the same stars, our mean abundances for thick and thin disc stars are consistent with APOGEE values. At the same time, both our and the APOGEE [Mg/Fe] trends pass through subsolar [Mg/Fe] at solar metallicity. Meanwhile, the [Mg/Fe] trend for warmer stars in the GILD sample pass through supersolar [Mg/Fe] values at solar metallicities with a systematic difference of $\sim$ 0.1 - 0.2 dex with respect to our [Mg/Fe] trend. In Figure~\ref{fig:mgindiv}, we, therefore, shifted the GILD trend down by 0.1 dex in order to normalize this comparison sample to the solar value, to allow for a  better comparison. We find that the LTE [Mg/Fe] values for our stars agree well with the original GILD trend (see Section~\ref{sec:LTEvsNLTE}). The systematic difference between our trend and GILD trend might be caused by the use of NLTE grids in this work and LTE values in GILD.

%While the difference in trends between different lines can be attributed to the uncertainties in the line data 

   %the line parameters we deAs mentioned in the section~\ref{sec:linelist}, we have used the best available line strength values from NIST or adjusted by astrophysical comparison and broadness parameters for our chosen set of lines

\subsection{Silicon (Si)}
\label{sec:si}

We determined [Si/Fe] for the stars in the nearby M giant sample and the solar neighborhood sample from one line at 16434.93 \AA\, in the H band and three lines in the K band: 20804.20 \AA, 20890.37 \AA,, and  20926.14 \AA. In Figure~\ref{fig:sispectra}, we plot the selected Si lines in the observed spectra of one thin disc star (top row panels) and one thick disc star (bottom row panels) in black circles, the line masks in yellow band, the synthetic spectrum fit to these lines in red, the synthetic spectrum without the Si I feature in green, and the variation in the fit resulting $\pm$0.2 dex change in the abundance value as red band. The stellar parameters estimated for each star are listed in the left most panels along with the derived abundance from each line in the corresponding panel. In Figure~\ref{fig:siindiv} we plot our individual [Si/Fe] trends in the top panels and the mean [Si/Fe] in the bottom panel (red circles and black star symbols for thin disc stars and orange diamonds for thick disk stars) along with [Si/Fe] from GILD sample (gray diamonds) and APOGEE DR17 (blue inverted triangles and cyan hexagons). 

 % At all metallicities, there is significant scatter and higher values for the [Si/Fe] abundances determined from the line at 21204.42 \AA.
 The synthetic spectrum fit to all four lines for the sample thin and thick disc stars in  Figure~\ref{fig:sispectra} are reasonably good. We find enhancements in the mean [Si/Fe] for all six thick disc stars. We note that there is only one good Si line for the solar metallicity thick disc star, $\#$4 ([Fe/H]=0.0), from which we determine high [Si/Fe] value of $\sim$ 0.18 dex. This could be the reason for the distinctly high [Si/Fe] but lower abundance values of other alpha elements for this star. The mean thin disc trend of the solar neighborhood and nearby M giant sample is consistent with the GILD sample thin disc trend at all metallicities, but we find $\sim$0.05 - 0.1 dex enhancement in our thick disc trend compared to the GILD thick disc trend. Similar difference is seen with respect to the APOGEE [Si/Fe] abundances of thick disc stars and also compared to the [Fe/H]<-0.5 thin disc stars. Overall, the thin-disc trend shows a small scatter and the thick-disk trend is clearly elevated in [Si/Fe].

\subsection{Calcium (Ca)}
\label{sec:ca}

We used five calcium absorption lines to determine [Ca/Fe], three in the H band: 16150.76 \AA, 16155.24 \AA, and 16157.36 \AA\, and two in the K band: 20962.57 \AA, and 20972.53 \AA. In the Figure~\ref{fig:caspectra}, we plot the selected Ca lines in the observed spectra of one thin disc star (top row panels) and one thick disc star (bottom row panels) in black circles, the line masks in yellow band, the synthetic spectrum fit to these lines in red, the synthetic spectrum without the Ca I feature in green, and the variation in the fit resulting $\pm$0.2 dex change in the abundance value as red band. The stellar parameters estimated for each star is listed in the left most panels along with the derived abundance from each line in the corresponding panel. Figure~\ref{fig:caindiv} shows the [Ca/Fe] trends from the six individual lines in the top panels and the mean [Ca/Fe] trend from the chosen set of good lines for each star from the solar neighborhood sample and six nearby M giants in the bottom panel.

As shown in the Figure~\ref{fig:caspectra}, we could fit all five lines very well. In addition, these lines are strongly dependent on the Ca abundance as indicated by the significant variation in the synthetic spectrum with $\pm$ 0.2 dex variation in [Ca/Fe]. The [Ca/Fe] trends from all six lines follow a downward trend, although with a shallower slope than for, e.g., [Mg/Fe].  Above solar metallicity [Ca/Fe] seem to at least level off to decrease again at [Fe/H] $>$ 0.2 dex. A larger sample would clarify this observation. The mean [Ca/Fe] values follow the same trend and the five metal poor thick disc stars have enhanced mean abundances compared to the thin disc stars. We do not find similar enhancement for the solar metallicity thick disc star. Unlike the case of Mg and Si, there is no clear separation between the thin and thick disc stars in the [Ca/Fe] trend from the GILD sample. Compared to the GILD trend, our mean [Ca/Fe] trend is systematically lower by 0.05 dex but pass through solar [Ca/Fe] at solar metallicity. Similar to Mg, we have, therefore, shifted the GILD trend down by 0.05 dex in order to normalize this comparison sample to the solar value, to allow for a  better comparison. The [Ca/Fe] trend from APOGEE DR17 sample follows a downward trend that continues to decrease at higher metallicities. The DR17 [Ca/Fe] values are lower for the thick disc stars and thus do not show any clear separation with respect to the thin disc stars.

\begin{figure*}
  \includegraphics[width=\textwidth]{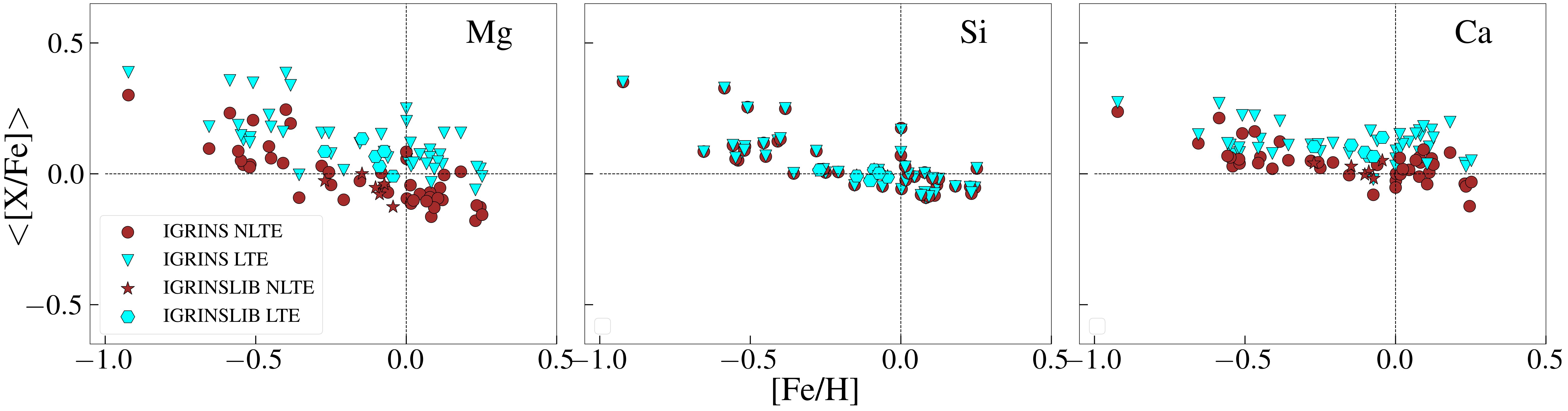}
  \caption{NLTE (brown) and LTE (cyan) abundance trends as a function of [Fe/H] for [Mg/Fe], [Si/Fe] and [Ca/Fe] estimated for 44 solar neighborhood M giants (circle symbol - NLTE and inverted triangle symbol - LTE) and six nearby M giants (star symbol - NLTE and hexagon symbol - LTE). There is negligible difference between NLTE and LTE measurements for Si while Mg and Ca show 0.05 - 0.15 dex differences. }
  \label{fig:lte-nlte}%
\end{figure*}

\subsection{Titanium (Ti)}
\label{sec:ti}

Only two lines have been used to determine [Ti/Fe], one in the H band: 16330.54 \AA\, and one in the K band: 21149.62 \AA. In Figure~\ref{fig:tispectra}, we plot the selected Ti lines in the observed spectra of one thin disc star (top row panels) and one thick disc star (bottom row panels) in black circles, the line masks in yellow band, the synthetic spectrum fit to these lines in red, the synthetic spectrum without the Ti I feature in green, and the variation in the fit resulting $\pm$0.2 dex change in the abundance value as red band. The stellar parameters estimated for each star is listed in the left most panels along with the derived abundance from each line in the corresponding panel. Similar to Mg, Si and Ca, Figure~\ref{fig:tiindiv} shows the [Ti/Fe] trends from the two individual lines in the top two panels and the mean [Ti/Fe] trend in the bottom panel.

Compared to the H band line at 16330.54 \AA, the K band line at 21149.62 \AA\, is weaker and blended with CN ($\nu$=1-3). The [Ti/Fe] trend determined using the line at 16330.54 \AA\, is found to have lower scatter at sub solar metallicities compared to the trend determined from the line at 21149.62 \AA. This may be attributed to the CN ($\nu$ = 1-3) blend in the left wing of the K band line as can be seen in Figure~\ref{fig:tispectra}. This scatter is carried forward to the mean [Ti/Fe] trend for low metallicity stars (\feh\,$<$ -0.4 dex). Yet the metal poor thick disc stars show clear enhancement compared to the thin disc stars. Furthermore, the mean [Ti/Fe] abundance trend is consistent with the optical [Ti/Fe] trend from the GILD sample. Unfortunately, APOGEE DR17 do not provide [Ti/Fe] measurements for the 44 stars. It has been noted that the Ti I lines in the APOGEE wavelength regime (H band) are very sensitive to \teff\, (\citealt{Jonsson:2018,Jonsson:2020}). Hence, it is reassuring for our method and especially for our determined \teff\, that the [Ti/Fe] determined using our stellar parameters are consistent with the trend from optical spectra.

\subsection{LTE and non-LTE comparison}
\label{sec:LTEvsNLTE}

As mentioned in the Section~\ref{sec:analysis}, we have applied NLTE corrections to the abundance measurements of Mg, Si and Ca. In addition, we estimated abundances for all three elements without applying NLTE corrections. Figure~\ref{fig:lte-nlte} shows the NLTE (brown) and LTE trends (cyan) for all three elements as a function of [Fe/H]. The NLTE-LTE difference for [Mg/Fe] and [Ca/Fe] range from -0.05 to -0.15 dex and there is no significant NLTE-LTE difference for [Si/Fe]. Thus applying NLTE corrections lower the abundance values for Mg and Si.

\section{Conclusions}
\label{sec:conclusion}

For abundance studies, a general method to determine the important stellar parameters for M giants is needed. For example, in order to study stellar abundances and the chemical evolution of stellar populations in dust obscured regions, like the inner bulge and Galactic Center region, the intrinsically bright M giants observed in the near-infrared is an optimal option. Not only is the extinction lower in the infrared, but cool stars, such as M giants, can only be analysed in the near-IR due to the ubiquitous TiO features in optical spectra of stars with \teff\, $<$ 4000\,K. Due to uncertainties in photometric methods, a method to determine the stellar parameters for M giants from the near-IR spectra themselves is needed. Such a method would thus open up the possibility to efficiently allow these regions to be analysed. 

Thus, in this quest to determine reliable stellar parameters and elemental abundances of M giant stars from their near infrared spectra, we have developed an iterative method, tested for  3400$\lesssim$ \teff $\lesssim$4000\,K, presented here. We have used high resolution (R$\sim$45,000) spectra in the H band observed with the IGRINS spectrometer.  IGRINS spectra cover the entire wavelength range of the H and K bands ($\sim$ 14 000 \AA to 25 000 \AA) which provides a wealth of spectral lines for the subsequent study of stellar abundances too.

%This method primarily relies on a selected set of $\sim$ 15 to 20 molecular OH lines to constrain \teff\,. To break the degeneracy between oxygen abundance and \teff\,   
 
Since \teff\, is one of the most crucial fundamental stellar parameter, we started off by identifying $\sim$ 15 to 20 \teff\, sensitive molecular OH lines in the H band based on our investigations using the APOGEE M giant spectra and their parameters provided in the DR17 catalog. In addition to \teff, the OH line strengths also depend on the oxygen abundance resulting in a degeneracy between \teff\, and oxygen abundance. Hence we fix the oxygen abundance based on the metallicity of the star following the [O/Fe] versus [Fe/H] trend in \cite{Amarsi:2019} for thin and thick disc stars (Figure~\ref{fig:Otrend}). 
It is reassuring that even for an initial misidentification of the stellar population, this will be able to be remedied by determining, e.g., the [Mg/Fe] abundance. This will clearly show which population the star actually belongs to (see Section~\ref{sec:uncertain}).

%To estimate the uncertainties in the derived parameters that could arise from the wrong assumption of stellar population (or wrong oxygen abundance), we determined them assuming thin disc for thick disc stars and vice versa. This led to $\sim$ $\pm$0.15-0.2 dex differences in [O/Fe]. 
%which in turn corresponds to differences of $\pm$ 100 K in \teff\,, $\pm$0.2 dex in \logg\,, $\pm$0.1 dex in \feh\,, $\pm$0.1 km/s in $\xi_\mathrm{micro}$, $\pm$0.1 dex in [C/Fe], and $\pm$0.1 dex in [N/Fe].   
%This in turn implies that for our method, the most important assumption is regarding the stellar population to which the star belongs: thin or thick disc (i.e. low oxygen or high oxygen abundance). We have shown that even for a wrong identification of the stellar population, the alpha trend subsequently determined will makes it 

After the initial assumption of the stellar population, we start with initial \teff\ and \feh\, of 3500 K and 0.0 dex, respectively. 
 %The oxygen abundance corresponding to the solar metallicity is fixed from the Amarsi trend. 
The \logg\, value is chosen to be 0.65 dex from the 10 Gyr Yonsei-Yale (YY) isochrones corresponding to the initial assumed \teff\, and \feh\, by means of simple linear interpolation \citep{rich:17}. Our method requires multiple iterations wherein firstly we determine \teff, \feh, $\xi_\mathrm{micro}$, [C/Fe] and [N/Fe] using SME by fitting the selected set of OH, CN, CO and Fe lines. We then determine \logg\, based on the newly determined \teff\, and \feh\, from the 10 Gyr YY isochrones. Similarly, the oxygen abundance is updated corresponding to the new \feh. With the updated \logg, and oxygen abundance, we determine new values of \teff, \feh, $\xi_\mathrm{micro}$, [C/Fe] and [N/Fe] and this cycle is repeated until there is negligible differences between the latest and previously determined values of all free parameters (Figure~\ref{fig:flow}). A final check of the assumed stellar population is then done based on the determined $\alpha$ abundances, most importantly the [Mg/Fe] abundance.

We validate our method by deriving the stellar parameters for six nearby well-studied M-giants with spectra from the IGRINS spectral library. Further, we demonstrate the accuracy and precision by determining the $\alpha$-element trends versus metallicity for 44 solar neighbourhood M giants from our two IGRINS runs on Gemini (We also include the six M giants in the $\alpha$-element trends versus metallicity). 
%which are also observed by APOGEE and compare with optical trends.%We tested our method using six nearby M giants in the IGRINS spectral library and 44 solar neighborhood M giants from our two IGRINS runs on Gemini. 
43 of the stars from our two IGRINS runs have also been analysed in APOGEE. The effective temperatures that we derive from our new method agree excellently with the six nearby, well-studied M giants (in Fig. \ref{fig:benchmark}) which indicates that the accuracy is indeed high. For the 43 solar neighborhood M giants, we find excellent agreement with APOGEE for \teff, \logg, \feh,  $\xi_\mathrm{micro}$, [C/Fe], [N/Fe], and [O/Fe] with mean differences and scatter (our method - APOGEE) of -67$\pm$33 K, -0.31$\pm$0.15 dex, 0.02$\pm$0.05 dex, 0.22$\pm$0.13 km/s, -0.05$\pm$0.06 dex, 0.06$\pm$0.06 dex, and 0.02$\pm$0.09 dex, respectively. Furthermore, the tight  offset with a small dispersion compared to APOGEE's \teff\ points to a high precision in both our derived temperatures and those derived from the APOGEE pipeline (Figure~\ref{fig:teffdiff}). The large scatter in the \teff\, determined using photometric methods like the IRFM and \teff-(V-K) relations further emphasizes the necessity of developing such spectroscopic methods to determine stellar parameters.

Typical uncertainties in the stellar parameters corresponding to $\pm$ 0.15 dex uncertainty in [O/Fe] are found to be $\pm$ 100 K in \teff\,, $\pm$0.2 dex in \logg\,, $\pm$0.1 dex in \feh\,, $\pm$0.1 km/s in $\xi_\mathrm{micro}$, $\pm$0.1 dex in [C/Fe], and $\pm$0.1 dex in [N/Fe]. Another source of uncertainty is the use of 10 Gyr YY isochrone to constrain \logg\,. However, we find that the difference in derived surface gravity is less than 0.1 dex on using a 2 Gyr YY isochrone instead of the 10 Gyr one. 

The $\alpha$-element trends versus metallicity for Mg, Si, Ca and Ti shows excellent agreement with both APOGEE DR17 trends for the same stars as well as with the GILD optical trends. We also find clear enhancement in abundances for thick disc stars. 

%The line-by-line scatter in the determined abundances for each star is listed in the Tables~\ref{table:mgfe} to ~\ref{table:tife}. The uncertainties in abundance determined from each line arising from stellar parameter uncertainties for seven selected stars are listed in the Table~\ref{table:uncert_params}.
%Summarioze the uncertainties: oxygen, isochrones, ... 

%Typical uncertainties in the stellar parameters we estimate to ... GOVIND...

The two main limitations of our method are fixing the oxygen abundance and the use of theoretical isochrones to constrain \logg. We explored in detail the effect of fixing wrong oxygen abundances on the derived stellar parameters, and found ways to remedy such wrong assumptions. To test the accuracy of the \logg\, values we estimated from isochrones in our method, it is imperative to test our method on  well-studied benchmark M giant stars \citep{Heiter:2015}. We also need to increase the sample size of the solar neighborhood M giants to cover full stellar parameter range, especially at temperatures lower than 3500 K.

As mentioned in the Section ~\ref{sec:intro}, \teff\,- line depth ratios (LDRs) relations have been proposed by many recent studies (\citealt{Fukue:2015}, \citealt{Taniguchi:2018}, \citealt{Taniguchi:2021}, \citealt{Matsunaga:2021}, \citealt{Afsar:2023}) especially in the near infrared wavelength regimes (0.97 - 2.4 $\mu$m) for late type giants with broad range of \teff s (3500 to 5400 K). Our method has only been tested on the cool stars (\teff<4000 K) and we need to apply our method on warmer stars with weaker molecular absorption lines in order to test its limits. At the same time, the outcome from the LDR methods are limited to \teff\, alone, while we determine the fundamental stellar parameters as well as C, N abundances with our method. 

To summarize, we have developed a method to determine reliable stellar parameters for M giants from their near infrared spectra, which can be  made fully automatic for future near-IR surveys. 

%We have tested best for 3500-4000K. 
%Caution: more benchmark star for lower temperatures. Need ot be tested. Caution. More test..
%Our method is well suited for 3500-4000K.
   
%Can be made fully automatic for future near-IR surveys. 

\begin{acknowledgements}
We thank the anonymous referee for the constructive comments and suggestions that improved the quality and aesthetics of the paper. G.N.\ acknowledges the support from the Wenner-Gren Foundations and the Royal Physiographic Society in Lund through the Stiftelsen Walter Gyllenbergs fond. G.N.\ thanks Henrik J\"onsson for enlightening discussions. N.R.\ acknowledge support from the Royal Physiographic Society in Lund through the Stiftelsen Walter Gyllenbergs fond and Märta och Erik Holmbergs donation and from Magnus Bergvalls stiftelse. This work used The Immersion Grating Infrared Spectrometer (IGRINS) was developed under a collaboration between the University of Texas at Austin and the Korea Astronomy and Space Science Institute (KASI) with the financial support of the US National Science Foundation under grants AST-1229522, AST-1702267 and AST-1908892, McDonald Observatory of the University of Texas at Austin, the Korean GMT Project of KASI, the Mt. Cuba Astronomical Foundation and Gemini Observatory. These results made use of the Lowell Discovery Telescope (LDT) at Lowell Observatory. Lowell is a private, non-profit institution dedicated to astrophysical research and public appreciation of astronomy and operates the LDT in partnership with Boston University, the University of Maryland, the University of Toledo, Northern Arizona University and Yale University. This paper includes data taken at The McDonald Observatory of The University of Texas at Austin.
\\
The following software and programming languages made this
research possible: TOPCAT (version 4.6; \citealt{topcat}); Python (version 3.8) and its packages ASTROPY (version 5.0; \citealt{astropy}), SCIPY \citep{scipy}, MATPLOTLIB \citep{matplotlib} and NUMPY \citep{numpy}.
\end{acknowledgements}



\bibliographystyle{aa}
\bibliography{references} % if your bibtex file is called example.bib

%%%%%%%%%%%%%%%%%%%%%%%%%%%%%%%%%%%%%%%%%%%%%%%%%%

%%%%%%%%%%%%%%%%% APPENDICES %%%%%%%%%%%%%%%%%%%%%

% \appendix
% \section{Table schema}

\begin{appendix} %First appendix

% \multirow{10}{*}{Mg} & 15740.716 $^{6}$   &  -0.440 $^{ *}$  &  1637 $^{3}$  &  0.280  \\
% & 15748.988 $^{6}$   &  -0.050 $^{ *}$  &  1637 $^{3}$  &  0.280  \\

% &  21204.492 $^{1 *}$   &  -0.388 $^{1 *}$  & -6.880 $^{1 }$   &    \\

 % &  16136.82 $^{1}$   &  -0.548 $^{1 *}$  &  1872 $^{3 }$  & 0.304   \\ 

\section{Tables}

\begin{table}[h]
\caption{Line parameters of Fe I lines used in this work. }\label{table:lines}
\begin{tabular}{l c c c c}
\hline
 Element & Wavelength  & log (gf) &  Broadening &  $\alpha$\\
 & ($\AA$)  & &   by H ${^a}$  &  \\
\hline
\multirow{30}{*}{Fe} 
 &  15485.454 $^{3}$   &  -0.828 $^{3 *}$  &  -7.290  &    \\ 
 &  15490.881 $^{3}$   &  0.789 $^{3}$  &  1618 $^{1}$  & 0.325   \\
 &  15500.799 $^{3}$   &  0.789 $^{3 *}$  &  1772 $^{1}$  & 0.320   \\
 &  15501.320 $^{3}$   &  0.789 $^{3 *}$  &  1620 $^{1}$  & 0.325   \\
 &  15502.174 $^{3}$   &  -1.034 $^{3 *}$  &  1876 $^{1}$  & 0.317   \\
 &  15662.013 $^{3}$   &  0.145 $^{3 *}$  &  1326 $^{2}$  & 0.235   \\
 &  15878.444 $^{3}$   &  -0.339 $^{3 *}$  &  -7.450  &    \\
 &  16153.247 $^{3}$   &  -0.682 $^{3 *}$  &  924 $^{2}$  & 0.229   \\
 &  16165.029 $^{3}$   &  0.782 $^{3 *}$  &  1598 $^{2}$  & 0.325   \\
 &  16171.930 $^{3}$   &  -0.399 $^{3 *}$  &  1858 $^{2}$  & 0.373   \\
 &  16174.975 $^{3}$   &  0.201 $^{3 *}$  &  1861 $^{1}$  & 0.316   \\
 &  16177.085 $^{3}$   &  -1.089 $^{3 *}$  &  -7.330  &    \\
 &  16179.583 $^{3}$   &  0.118 $^{3 *}$  &  1600 $^{1}$  & 0.325   \\
 &  16180.900 $^{3}$   &  0.203 $^{3 *}$  &  1431 $^{1}$  & 0.329   \\
 &  16182.170 $^{3}$   &  -0.805 $^{3 *}$  &  1599 $^{1}$  & 0.325   \\
 &  16185.799 $^{3}$   &  0.211 $^{3 *}$  &  1919 $^{1}$  & 0.313   \\
 &  16245.763 $^{3}$   &  -0.652 $^{3 *}$  &  1602 $^{1}$  & 0.325   \\
 &  16246.458 $^{4}$   &  -0.106 $^{3 *}$  &  1419 $^{1}$  & 0.329   \\
 &  16258.912 $^{3}$   &  -0.825 $^{3 *}$  &  -7.290  &    \\
 &  16316.320 $^{3}$   &  0.887 $^{3 *}$  &  1424 $^{1}$  & 0.329   \\
 &  16318.690 $^{3}$   &  -0.436 $^{3 *}$  &  -7.550  &    \\
 &  16324.451 $^{3}$   &  -0.522 $^{3 *}$  &  936 $^{1}$  & 0.229   \\
 &  16331.524 $^{3}$   &  -0.506 $^{3 *}$  &  -7.470  &    \\
 &  16333.141 $^{3}$   &  -1.356 $^{3 *}$  &  -7.450  &    \\
 &  16517.223 $^{3}$   &  0.572 $^{3 *}$  &  1414 $^{1}$  & 0.329   \\
 &  16612.761 $^{3}$   &  0.018 $^{3 *}$  &  1846 $^{1}$  & 0.315   \\
 &  16619.737 $^{3}$   &  -1.524 $^{3 *}$  &  1082 $^{1}$  & 0.228   \\
 &  16783.037 $^{3}$   &  -0.695 $^{3 *}$  &  1409 $^{1}$  & 0.329   \\
 &  16792.224 $^{3}$   &  -0.890 $^{3 *}$  &  1588 $^{1}$  & 0.324   \\
 &  16794.210 $^{3}$   &  -0.362 $^{3 *}$  &  -7.430  &    \\
 
\hline 
\hline
\end{tabular}
\tablefoot{a: Collisional broadening by neutral hydrogen gives either the broadening cross section $\sigma(v=10^4\mbox{m/s})$ in atomic units ($a_0^2$), with velocity parameter $\alpha$ (see text for more details), or if the value is negative and no $\alpha$ given, then the line width is given in the standard form  $\log \Gamma/N_H$ at $T=10000$~K, where $\Gamma$ is the full-width at half maximum in $\mbox{rad}/\mbox{s}$ and $N_H$ is in cm$^{-3}$. 1: BSYN based on routines from MARCS code \cite{marcs:08}, 2: Paul Barklem (private communication), 3: \cite{K14}, 4: \cite{Biemont:1985}, *: astrophysical estimate }
\end{table}

 \begin{table}[b]
\caption{Line parameters of the lines of each element from which abundances have been derived. }\label{table:lines}
\begin{tabular}{l c c c c}
\hline
 Element & Wavelength  & log (gf) &  Broadening &  $\alpha$\\
 & ($\AA$)  & &   by H ${^a}$  &  \\
\hline
\multirow{8}{*}{Mg} & 21059.757 $^{4 *}$   &  -0.384 $^{5 *}$  &  4440 $^{3}$  &  1.10  \\
&  21060.710 $^{4 *}$   & -0.530 $^{5 }$   & 4440 $^{3}$   &   1.10  \\
&  21060.896 $^{4 *}$   & -1.587 $^{5 }$   & 4440 $^{3}$   &    1.10 \\
&  21060.896 $^{4 *}$   & -0.407 $^{5 }$   & 4440 $^{3}$   &   1.10  \\
&  21061.095 $^{4 *}$   & -3.383 $^{5 }$   & 4440 $^{3}$   &   1.10  \\
&  21061.095 $^{4 *}$   & -1.583 $^{5 }$   & 4440 $^{3}$   &    1.10 \\
&  21061.095 $^{4 *}$   & -0.298 $^{5 }$   & 4440 $^{3}$   &   1.10  \\
&  21458.962 $^{6}$   & -1.319 $^{6}$   & 4840 $^{3}$   &   0.359  \\
\hline
\multirow{5}{*}{Si} & 16434.933 $^{1 *}$   & -1.483 $^{1 *}$   & 862 $^{3}$   &  0.288  \\
 & 20804.225 $^{1 *}$   & -1.026 $^{1 *}$   & 861 $^{2}$   &  0.292  \\
 &  20890.415 $^{1 *}$   & -1.613 $^{1 *}$   & 859 $^{2}$   &  0.292  \\
 &  20926.149 $^{1}$   &  -1.076 $^{1 *}$  &  1484 $^{2}$ & 0.324     \\
\hline
\multirow{5}{*}{Ca}
 &  16150.75 $^{1 *}$   &  -0.244 $^{1 *}$  &  1872 $^{3 }$  & 0.304   \\
 &  16155.244 $^{1 *}$   &  -0.674 $^{1 *}$  &  1872 $^{3 }$  & 0.304   \\
 &  16157.356 $^{1 *}$   &  -0.214 $^{1 *}$  &  1946 $^{3 }$  & 0.284   \\
 &  20962.570 $^{1 *}$   &  -0.784 $^{1 *}$  &  -7.230 $^{1 }$  &    \\ 
 &  20972.529 $^{1 }$   &  -1.002 $^{1 *}$  &  -7.230 $^{1 }$  &    \\ 
\hline
\multirow{2}{*}{Ti} 
 &  16330.532 $^{7}$   &  -0.906 $^{7 *}$  &  484 $^{2}$  & 0.239   \\ 
 &  21149.625 $^{8 *}$   &  0.789 $^{8 *}$  &  -7.440  &    \\
 \hline
\hline
\end{tabular}
\tablefoot{a: Collisional broadening by neutral hydrogen gives either the broadening cross section $\sigma(v=10^4\mbox{m/s})$ in atomic units ($a_0^2$), with velocity parameter $\alpha$ (see text for more details), or if the value is negative and no $\alpha$ given, then the line width is given in the standard form  $\log \Gamma/N_H$ at $T=10000$~K, where $\Gamma$ is the full-width at half maximum in $\mbox{rad}/\mbox{s}$ and $N_H$ is in cm$^{-3}$. 1: \cite{K07}, 2: BSYN based on routines from MARCS code \cite{marcs:08}, 3: Paul Barklem (private communication), 4: \cite{Brault:1983}, 5: \cite{civis:2013}, 6: manual entry by NIST lookup, 7: \cite{LGWSC}, 8: \cite{K16}, *: astrophysical estimate }
\end{table}

\begin{table}
\caption{Spectral line data of the OH molecular lines used in the method to determine stellar parameters. }\label{table:OH}
\begin{tabular}{l c c}
\hline
 Molecule & Wavelength  & log (gf) \\
 & ($\AA$)  &   \\
\hline
\multirow{27}{*}{OH } 
& 15236.623   &  -5.861     \\
& 15236.956   &  -5.861     \\
& 15391.057   &  -5.512     \\
& 15391.205   &  -5.512     \\
& 15469.762   &  -5.242     \\
& 15470.216   &  -5.242     \\
& 15505.324   &  -5.378     \\
& 15505.746   &  -5.378     \\
& 15651.897   &  -5.203     \\
& 15653.480   &  -5.203     \\
& 15654.112   &  -6.808     \\
& 16247.884   &  -5.177     \\
& 16312.494   &  -5.077     \\
& 16312.920   &  -5.077     \\
& 16346.182   &  -5.002     \\
& 16347.493   &  -5.002     \\
& 16352.213   &  -4.897     \\
& 16581.269   &  -4.874     \\
& 16714.359   &  -4.758     \\
& 16872.277   &  -5.032     \\
& 16886.275   &  -4.874     \\
& 16886.293   &  -6.976     \\
& 16895.183   &  -4.743     \\
& 16895.319   &  -5.706     \\
& 16898.778   &  -4.743     \\
& 16902.733   &  -4.674     \\
& 16909.289   &  -4.712     \\
\hline 
\hline
\end{tabular}

\end{table}

% \begin{table}
% \caption{Spectral line data for the lines used }\label{table:lines}
% \begin{tabular}{l c c}
% \hline
%  Molecule & Wavelength  & log (gf) \\
%  & ($\AA$)  &   \\
% \hline
% \multirow{15}{*}{OH ($\nu$=3-1)} 
% & 15236.623   &  -5.861     \\
% & 15236.956   &  -5.861     \\
% & 15391.057   &  -5.512     \\
% & 15391.205   &  -5.512     \\
% & 15505.324   &  -5.378     \\
% & 15505.746   &  -5.378     \\
% & 15651.897   &  -5.203     \\
% & 15653.480   &  -5.203     \\
% & 15654.112   &  -6.808     \\
% & 16352.213   &  -4.897     \\
% & 16714.359   &  -4.758     \\
% & 16895.183   &  -4.743     \\
% & 16895.319   &  -5.706     \\
% & 16898.778   &  -4.743     \\
% & 16909.289   &  -4.712     \\
% \hline
% \multirow{10}{*}{OH ($\nu$=4-2)} 
% & 15469.762   &  -5.242     \\
% & 15470.216   &  -5.242     \\
% & 16312.494   &  -5.077     \\
% & 16312.920   &  -5.077     \\
% & 16346.182   &  -5.002     \\
% & 16347.493   &  -5.002     \\
% & 16581.269   &  -4.874     \\
% & 16886.275   &  -4.874     \\
% & 16886.293   &  -6.976     \\
% & 16902.733   &  -4.674     \\
% \hline
% \multirow{2}{*}{OH ($\nu$=2-0)} 
% & 16247.884   &  -5.177     \\
% & 16872.277   &  -5.032     \\
% \hline 
% \hline
% \end{tabular}

% \end{table}

\begin{table}[h]
\caption{Spectral line data of the CN molecular lines used in the method to determine stellar parameters }\label{table:CN}
\begin{tabular}{l c c}
\hline
 Molecule & Wavelength  & log (gf) \\
 & ($\AA$)  &   \\
\hline
\multirow{25}{*}{CN} 
& 15466.235   &  -1.195     \\
& 15471.812   &  -1.749     \\
& 15485.339   &  -1.779     \\
& 15489.56   &  -1.927     \\
& 15489.764   &  -1.932     \\
& 15494.748   &  -1.542     \\
& 15495.256   &  -1.196     \\
& 15496.319   &  -3.095     \\
& 15500.927   &  -1.742     \\
& 15501.498   &  -1.764     \\
& 15660.700   &  -1.521     \\
& 15661.595   &  -1.56     \\
& 15871.453   &  -1.633     \\
& 15881.108   &  -1.847     \\
& 16021.961   &  -1.518     \\
& 16179.943   &  -1.619     \\
& 16180.109   &  -1.131     \\
& 16329.245   &  -1.553     \\
& 16334.003   &  -1.932     \\
& 16352.092   &  -1.600     \\
& 16581.908   &  -1.215     \\
& 16582.026   &  -1.526     \\
& 16582.199   &  -1.109     \\
& 16615.321   &  -1.474     \\
& 16618.553   &  -1.891     \\
& 16791.783   &  -1.870     \\
& 16895.399   &  -1.462     \\
\hline 
\hline
\end{tabular}

\end{table}

\begin{table}[b]
\caption{Spectral line data of the CO molecular lines used in the method to determine stellar parameters }\label{table:CO}
\begin{tabular}{l c c}
\hline
 Molecule & Wavelength  & log (gf) \\
 & ($\AA$)  &   \\
\hline
\multirow{43}{*}{CO} 
& 16025.53   &  -6.275     \\
& 16025.935   &  -7.004     \\
& 16026.94   &  -7.851     \\
& 16030.941   &  -6.262     \\
& 16031.274   &  -7.038     \\
& 16184.507   &  -6.26     \\
& 16184.565   &  -6.242     \\
& 16184.738   &  -6.279     \\
& 16184.912   &  -6.224     \\
& 16185.256   &  -6.298    \\
& 16185.548   &  -6.207     \\
& 16186.062   &  -6.317     \\
& 16186.473   &  -6.19     \\
& 16187.154   &  -6.337     \\
& 16187.688   &  -6.173     \\
& 16332.817   &  -5.79      \\
& 16334.087   &  -7.661     \\
& 16351.905   &  -7.148     \\
& 16352.478   &  -5.766     \\
& 16613.174   &  -5.777     \\
& 16613.225   &  -5.795     \\
& 16613.427   &  -5.759     \\
& 16613.579   &  -5.814     \\
& 16613.984   &  -5.742     \\
& 16614.236   &  -5.833     \\
& 16614.846   &  -5.724     \\
& 16615.195   &  -5.853     \\
& 16616.012   &  -5.708     \\
& 16616.457   &  -5.873     \\
& 16617.485   &  -5.691     \\
& 16618.021   &  -5.893     \\
& 16619.264   &  -5.675     \\
& 16619.887   &  -5.914     \\
& 16620.177   &  -7.028     \\
& 16871.668   &  -5.944     \\
& 16885.696   &  -6.849     \\
& 16895.397   &  -6.434     \\
& 16896.04   &  -5.293     \\
& 16899.304   &  -6.11     \\
& 16901.802   &  -6.527     \\
& 16902.455   &  -5.28     \\
& 16909.199   &  -5.266     \\
& 16909.543   &  -6.843     \\
\hline 
\hline
\end{tabular}

\end{table}

\begin{table*}
\caption{  }\label{table:mgfe}
\begin{tabular}{c c c c c c c}
\hline
\hline
 Index / Star & 21059.76 \AA  & 21060.89 \AA  &  21458.87 \AA   & $<$[Mg/Fe]$>$ & $\sigma$[Mg/Fe] / $\sqrt{N_{lines}}$  \\
\hline
1  &  -0.01  &  0.02  &  0.08  &  0.03  &  0.02  \\ 
2  &  0.03  &  0.06  &  0.09  &  0.06  &  0.01  \\ 
3  &  0.21  &  0.14  &  0.26  &  0.21  &  0.02  \\ 
4  &  0.07  &  0.02  &  0.08  &  0.06  &  0.01  \\ 
5  &  -0.07  &  -0.04  &  -0.03  &  -0.05  &  0.01  \\ 
6  &  -0.01  &  0.0  &  0.12  &  0.04  &  0.02  \\ 
7  &  0.06  &  0.07  &  0.16  &  0.1  &  0.02  \\ 
8  &  0.21  &  0.17  &  0.2  &  0.19  &  0.01  \\ 
9  &  0.04  &  -0.01  &  0.07  &  0.04  &  0.02  \\ 
10  &  0.26  &  0.3  &  0.34  &  0.3  &  0.02  \\ 
11  &  0.01  &  -0.01  &  0.08  &  0.03  &  0.02  \\ 
12  &  0.05  &  -0.0  &  0.1  &  0.05  &  0.02  \\ 
13  &  -0.16  &  -0.08  &  -0.06  &  -0.1  &  0.02  \\ 
14  &  0.07  &  0.05  &  0.14  &  0.09  &  0.02  \\ 
15  &  -0.14  &  -0.12  &  -0.02  &  -0.09  &  0.02  \\ 
16  &  -0.18  &  -0.11  &  -0.09  &  -0.13  &  0.02  \\ 
17  &  0.04  &  0.09  &  0.18  &  0.1  &  0.03  \\ 
18  &  -0.07  &  0.07  &  0.12  &  0.04  &  0.03  \\ 
19  &  -0.06  &  -0.03  &  0.01  &  -0.03  &  0.01  \\ 
20  &  -0.03  &  -0.01  &  0.06  &  0.0  &  0.02  \\ 
21  &  -0.15  &  -0.09  &  -0.06  &  -0.1  &  0.02  \\ 
22  &  -0.12  &  -0.01  &  -0.04  &  -0.05  &  0.02  \\ 
23  &  0.08  &  -0.1  &  0.02  &  -0.0  &  0.03  \\ 
24  &  -0.05  &  -0.11  &  -0.05  &  -0.07  &  0.01  \\ 
25  &  -0.16  &  -0.2  &  -0.12  &  -0.16  &  0.02  \\ 26  &  -0.13  &  -0.07  &  -0.02  &  -0.08  &  0.02  \\ 
27  &  -0.12  &  -0.1  &  -0.07  &  -0.09  &  0.01  \\ 
28  &  -0.08  &  -0.11  &  -0.02  &  -0.07  &  0.02  \\ 
29  &  0.03  &  0.09  &  0.13  &  0.08  &  0.02  \\ 
30  &  -0.07  &  -0.06  &  0.0  &  -0.04  &  0.02  \\ 
31  &  -0.27  &  -0.15  &  -0.11  &  -0.18  &  0.03  \\ 
32  &  -0.12  &  -0.14  &  -0.07  &  -0.11  &  0.01  \\ 
33  &  -0.12  &  -0.12  &  -0.06  &  -0.1  &  0.01  \\ 
34  &  -0.01  &  0.0  &  0.03  &  0.01  &  0.01  \\ 
35  &  -0.02  &  0.01  &  0.03  &  0.01  &  0.01  \\ 
36  &  -0.11  &  -0.1  &  -0.05  &  -0.09  &  0.01  \\ 
37  &  -0.12  &  -0.09  &  -0.07  &  -0.09  &  0.01  \\ 
38  &  -0.18  &  -0.13  &  -0.06  &  -0.12  &  0.02  \\ 
39  &  0.21  &  0.28  &  0.25  &  0.25  &  0.01  \\ 
40  &  -0.2  &  -0.13  &  -0.14  &  -0.16  &  0.01  \\ 
41  &  -0.18  &  -0.18  &  -0.03  &  -0.13  &  0.03  \\ 
42  &  -0.07  &  -0.07  &  0.01  &  -0.04  &  0.02  \\ 
43  &  -0.14  &  -0.12  &  -0.05  &  -0.1  &  0.02  \\ 
44  &  0.2  &  0.19  &  0.3  &  0.23  &  0.02  \\ 
HD132813  &  -0.04  &  -0.09  &  0.06  &  -0.03  &  0.03  \\ 
HD175588  &  -0.18  &  -0.19  &  -0.01  &  -0.13  &  0.03  \\ 
HD89758  &  -0.1  &  -0.13  &  -0.0  &  -0.08  &  0.02  \\ 
HD224935  &  -0.12  &  -0.1  &  0.06  &  -0.05  &  0.03  \\ 
HD101153  &  -0.11  &  -0.09  &  0.09  &  -0.04  &  0.04  \\ 
HIP54396  &  -0.06  &  -0.02  &  0.08  &  0.0  &  0.03  \\ 

\hline
\hline
\end{tabular}
\end{table*}

\begin{table*}
\caption{  }\label{table:sife}
\begin{tabular}{c c c c c c c c}
\hline
\hline
 Index / Star  & 16434.93 \AA  & 20804.20 \AA & 20890.37 \AA  & 20926.14 \AA  & $<$[Si/Fe]$>$ & $\sigma$[Si/Fe] / $\sqrt{N_{lines}}$  \\
\hline
1  &  0.21  &  0.04  &  0.08  &  0.02  &  0.09  &  0.03  \\ 
2  &  0.06  &  0.05  &  0.1  &  0.06  &  0.07  &  0.0  \\ 
3  &  0.37  &  0.28  &  0.17  &  0.2  &  0.26  &  0.04  \\ 
4  &  0.18  &  --  &  --  &  --  &  0.18  &  0.0  \\ 
5  &  0.06  &  0.06  &  0.09  &  -0.17  &  0.01  &  0.04  \\ 
6  &  0.15  &  0.04  &  0.02  &  -0.0  &  0.05  &  0.02  \\ 
7  &  0.11  &  --  &  0.06  &  --  &  0.08  &  0.01  \\ 
8  &  0.26  &  0.34  &  0.06  &  0.34  &  0.25  &  0.04  \\ 
9  &  0.1  &  0.13  &  0.13  &  0.06  &  0.1  &  0.02  \\ 
10  &  0.38  &  0.29  &  0.36  &  0.36  &  0.35  &  0.01  \\ 
11  &  0.1  &  0.03  &  -0.01  &  0.24  &  0.09  &  0.04  \\ 
12  &  0.03  &  0.13  &  0.07  &  0.0  &  0.06  &  0.02  \\ 
13  &  0.04  &  -0.07  &  0.05  &  0.0  &  0.01  &  0.02  \\ 
14  &  0.17  &  0.13  &  0.15  &  -0.01  &  0.11  &  0.02  \\ 
15  &  0.07  &  0.07  &  --  &  -0.13  &  0.0  &  0.04  \\ 
16  &  0.01  &  -0.06  &  -0.0  &  -0.15  &  -0.05  &  0.03  \\ 
17  &  0.22  &  --  &  0.02  &  --  &  0.12  &  0.05  \\ 
18  &  0.15  &  0.1  &  --  &  --  &  0.12  &  0.01  \\ 
19  &  -0.0  &  0.02  &  -0.12  &  -0.07  &  -0.04  &  0.02  \\ 
20  &  -0.03  &  -0.02  &  0.13  &  -0.08  &  -0.0  &  0.03  \\ 
21  &  -0.08  &  0.0  &  -0.02  &  -0.07  &  -0.04  &  0.02  \\ 
22  &  -0.01  &  -0.08  &  -0.11  &  -0.13  &  -0.08  &  0.02  \\ 
23  &  -0.04  &  -0.18  &  -0.08  &  0.22  &  -0.02  &  0.06  \\ 
24  &  -0.14  &  -0.01  &  --  &  -0.06  &  -0.07  &  0.02  \\ 
25  &  -0.08  &  -0.02  &  -0.22  &  -0.04  &  -0.09  &  0.03  \\ 
26  &  -0.16  &  0.08  &  --  &  0.05  &  -0.01  &  0.05  \\ 
27  &  -0.13  &  -0.01  &  -0.09  &  -0.0  &  -0.06  &  0.02  \\ 
28  &  0.02  &  -0.05  &  --  &  -0.11  &  -0.05  &  0.02  \\ 
29  &  0.12  &  0.07  &  0.02  &  0.07  &  0.07  &  0.02  \\ 
30  &  0.01  &  0.04  &  -0.05  &  0.02  &  0.01  &  0.01  \\ 
31  &  -0.06  &  -0.08  &  0.06  &  -0.11  &  -0.05  &  0.02  \\ 
32  &  -0.03  &  0.0  &  -0.08  &  0.01  &  -0.02  &  0.02  \\ 
33  &  -0.08  &  0.12  &  -0.12  &  0.08  &  0.0  &  0.05  \\ 
34  &  -0.16  &  0.01  &  -0.0  &  -0.03  &  -0.05  &  0.02  \\ 
35  &  0.04  &  0.03  &  0.0  &  0.0  &  0.02  &  0.0  \\ 
36  &  0.04  &  -0.0  &  0.01  &  -0.02  &  0.01  &  0.01  \\ 
37  &  0.04  &  0.01  &  0.04  &  -0.16  &  -0.02  &  0.03  \\ 
38  &  -0.18  &  -0.03  &  --  &  -0.02  &  -0.08  &  0.03  \\ 
39  &  0.19  &  0.05  &  0.12  &  0.18  &  0.13  &  0.02  \\ 
40  &  0.04  &  0.06  &  -0.06  &  0.04  &  0.02  &  0.02  \\ 
41  &  -0.14  &  -0.03  &  -0.14  &  -0.03  &  -0.09  &  0.03  \\ 
42  &  -0.0  &  0.11  &  -0.01  &  0.01  &  0.03  &  0.02  \\ 
43  &  -0.13  &  0.0  &  -0.11  &  -0.09  &  -0.08  &  0.02  \\ 
44  &  0.45  &  0.24  &  0.34  &  0.28  &  0.33  &  0.04  \\ 
HD132813  &  -0.05  &  0.07  &  --  &  --  &  0.01  &  0.03  \\ 
HD175588  &  -0.09  &  -0.0  &  0.12  &  -0.07  &  -0.01  &  0.04  \\ 
HD89758  &  0.1  &  0.01  &  -0.04  &  -0.01  &  0.02  &  0.02  \\ 
HD224935  &  -0.05  &  -0.11  &  0.01  &  0.06  &  -0.02  &  0.03  \\ 
HD101153  &  -0.02  &  -0.02  &  0.04  &  0.01  &  0.0  &  0.01  \\ 
HIP54396  &  -0.14  &  0.03  &  0.09  &  --  &  -0.01  &  0.05  \\ 

\hline
\hline
\end{tabular}
\end{table*}

\begin{table*}
\caption{  }\label{table:cafe}
\begin{tabular}{c c c c c c c c c}
\hline
\hline
 Index / Star  & 16136.82 \AA  & 16150.76 \AA & 16155.24 \AA  & 16157.36 \AA  &  20962.57 \AA  &  20972.53 \AA   & $<$[Ca/Fe]$>$ & $\sigma$[Ca/Fe] / $\sqrt{N_{lines}}$  \\
\hline
1  &  -0.03  &  -0.01  &  0.09  &  0.01  &  0.07  &  0.09  &  0.04  &  0.02  \\ 
2  &  -0.02  &  0.05  &  0.07  &  0.08  &  0.05  &  0.07  &  0.05  &  0.01  \\ 
3  &  0.14  &  0.14  &  0.18  &  0.11  &  0.16  &  0.19  &  0.15  &  0.01  \\ 
4  &  -0.13  &  -0.0  &  -0.02  &  -0.04  &  0.01  &  -0.08  &  -0.04  &  0.02  \\ 
5  &  -0.22  &  -0.18  &  0.0  &  -0.13  &  -0.01  &  -0.08  &  -0.1  &  0.04  \\ 
6  &  -0.07  &  0.01  &  0.04  &  -0.0  &  0.06  &  0.04  &  0.01  &  0.01  \\ 
7  &  0.09  &  0.06  &  0.13  &  0.01  &  0.17  &  0.21  &  0.11  &  0.02  \\ 
8  &  0.02  &  0.05  &  0.14  &  0.04  &  0.22  &  0.17  &  0.11  &  0.03  \\ 
9  &  0.02  &  -0.04  &  0.1  &  -0.02  &  0.1  &  0.07  &  0.04  &  0.02  \\ 
10  &  0.26  &  0.22  &  0.24  &  0.19  &  0.21  &  0.32  &  0.24  &  0.01  \\ 
11  &  0.01  &  0.04  &  0.08  &  0.01  &  0.03  &  0.13  &  0.05  &  0.02  \\ 
12  &  0.02  &  0.06  &  0.09  &  0.01  &  0.09  &  0.07  &  0.06  &  0.02  \\ 
13  &  -0.05  &  -0.01  &  0.07  &  -0.01  &  0.08  &  0.09  &  0.03  &  0.02  \\ 
14  &  0.01  &  0.0  &  0.1  &  0.02  &  0.11  &  0.11  &  0.06  &  0.02  \\ 
15  &  -0.04  &  -0.0  &  0.09  &  0.01  &  0.09  &  0.08  &  0.04  &  0.02  \\ 
16  &  -0.23  &  -0.18  &  -0.11  &  -0.17  &  -0.06  &  -0.09  &  -0.14  &  0.02  \\ 
17  &  0.05  &  -0.03  &  0.04  &  0.04  &  0.13  &  0.03  &  0.04  &  0.01  \\ 
18  &  -0.05  &  -0.01  &  0.03  &  -0.0  &  0.03  &  0.05  &  0.01  &  0.01  \\ 
19  &  -0.14  &  -0.09  &  0.04  &  -0.03  &  -0.0  &  0.06  &  -0.03  &  0.03  \\ 
20  &  -0.07  &  -0.01  &  0.1  &  0.02  &  0.12  &  0.1  &  0.04  &  0.03  \\ 
21  &  -0.06  &  -0.02  &  0.12  &  -0.02  &  0.11  &  0.11  &  0.04  &  0.03  \\ 
22  &  -0.15  &  -0.1  &  0.04  &  -0.06  &  0.07  &  0.08  &  -0.02  &  0.04  \\ 
23  &  -0.1  &  0.13  &  0.02  &  0.0  &  0.01  &  0.02  &  0.01  &  0.01  \\ 
24  &  -0.07  &  -0.06  &  0.03  &  -0.04  &  -0.02  &  0.03  &  -0.02  &  0.02  \\ 
25  &  -0.07  &  0.04  &  0.09  &  -0.04  &  0.09  &  0.04  &  0.02  &  0.03  \\ 
26  &  -0.08  &  -0.05  &  0.01  &  -0.01  &  0.06  &  0.07  &  0.0  &  0.02  \\ 
27  &  -0.11  &  -0.04  &  0.04  &  -0.07  &  0.05  &  0.02  &  -0.02  &  0.02  \\ 
28  &  -0.09  &  -0.02  &  0.04  &  -0.05  &  0.07  &  0.13  &  0.01  &  0.03  \\ 
29  &  -0.21  &  -0.17  &  -0.02  &  -0.08  &  0.01  &  -0.0  &  -0.08  &  0.04  \\ 
30  &  -0.06  &  -0.12  &  0.05  &  -0.02  &  0.09  &  0.12  &  0.01  &  0.03  \\ 
31  &  -0.18  &  -0.11  &  -0.0  &  -0.14  &  0.02  &  0.04  &  -0.06  &  0.04  \\ 
32  &  -0.1  &  -0.06  &  0.03  &  -0.05  &  0.03  &  0.05  &  -0.02  &  0.02  \\ 
33  &  -0.12  &  -0.01  &  0.07  &  -0.05  &  0.02  &  0.05  &  -0.01  &  0.02  \\ 
34  &  -0.07  &  -0.0  &  0.12  &  0.02  &  0.11  &  0.15  &  0.06  &  0.03  \\ 
35  &  -0.09  &  -0.04  &  0.08  &  -0.01  &  0.08  &  0.11  &  0.02  &  0.03  \\ 
36  &  -0.01  &  0.0  &  0.12  &  -0.03  &  0.06  &  0.16  &  0.05  &  0.03  \\ 
37  &  -0.16  &  -0.12  &  0.0  &  -0.09  &  0.0  &  0.01  &  -0.06  &  0.03  \\ 
38  &  -0.16  &  -0.12  &  -0.06  &  -0.12  &  0.03  &  0.03  &  -0.07  &  0.03  \\ 
39  &  0.12  &  0.07  &  0.19  &  0.14  &  0.24  &  0.17  &  0.16  &  0.02  \\ 
40  &  -0.13  &  -0.12  &  0.02  &  -0.15  &  0.01  &  0.09  &  -0.05  &  0.03  \\ 
41  &  -0.04  &  0.05  &  0.17  &  0.02  &  0.08  &  0.14  &  0.07  &  0.03  \\ 
42  &  0.01  &  0.02  &  0.12  &  -0.0  &  0.05  &  0.12  &  0.05  &  0.02  \\ 
43  &  -0.04  &  -0.0  &  0.1  &  -0.01  &  0.07  &  0.1  &  0.04  &  0.02  \\ 
44  &  0.21  &  0.16  &  0.24  &  0.14  &  0.26  &  0.27  &  0.21  &  0.02  \\ 
HD132813  &  -0.03  &  -0.02  &  0.02  &  -0.1  &  0.12  &  0.19  &  0.03  &  0.04  \\ 
HD175588  &  -0.06  &  -0.03  &  0.07  &  -0.1  &  0.13  &  0.18  &  0.03  &  0.04  \\ 
HD89758  &  -0.1  &  -0.08  &  0.06  &  -0.06  &  0.09  &  0.04  &  -0.01  &  0.03  \\ 
HD224935  &  -0.11  &  -0.07  &  0.04  &  -0.1  &  --  &  0.12  &  -0.02  &  0.04  \\ 
HD101153  &  -0.14  &  -0.06  &  0.02  &  -0.15  &  --  &  0.13  &  -0.04  &  0.04  \\ 
HIP54396  &  -0.11  &  -0.05  &  0.03  &  -0.09  &  0.13  &  0.13  &  0.01  &  0.04  \\ 

\hline
\hline
\end{tabular}
\end{table*}

\begin{table*}
\caption{  }\label{table:tife}
\begin{tabular}{c c c c c}
\hline
\hline
 Index / Star  & 16330.54 \AA  & 21149.62 \AA & $<$[Ti/Fe]$>$ & $\sigma$[Ti/Fe] / $\sqrt{N_{lines}}$  \\
\hline
1  &  0.08  &  0.11  &  0.1  &  0.01  \\ 
2  &  0.01  &  0.32  &  0.17  &  0.07  \\ 
3  &  0.2  &  0.31  &  0.25  &  0.03  \\ 
4  &  0.04  &  0.04  &  0.04  &  0.0  \\ 
5  &  -0.08  &  0.07  &  -0.01  &  0.04  \\ 
6  &  0.07  &  --  &  0.07  &  0.0  \\ 
7  &  0.13  &  0.31  &  0.22  &  0.04  \\ 
8  &  0.22  &  0.38  &  0.3  &  0.04  \\ 
9  &  0.09  &  -0.03  &  0.03  &  0.03  \\ 
10  &  0.36  &  0.43  &  0.39  &  0.01  \\ 
11  &  0.1  &  0.21  &  0.16  &  0.03  \\ 
12  &  0.08  &  0.2  &  0.14  &  0.03  \\ 
13  &  0.02  &  0.23  &  0.12  &  0.05  \\ 
14  &  0.12  &  0.34  &  0.23  &  0.05  \\ 
15  &  -0.03  &  0.23  &  0.1  &  0.06  \\ 
16  &  -0.1  &  0.01  &  -0.04  &  0.03  \\ 
17  &  0.04  &  0.06  &  0.05  &  0.01  \\ 
18  &  0.08  &  0.08  &  0.08  &  0.0  \\ 
19  &  0.01  &  0.01  &  0.01  &  0.0  \\ 
20  &  0.0  &  0.03  &  0.02  &  0.01  \\ 
21  &  0.0  &  0.02  &  0.01  &  0.01  \\ 
22  &  -0.09  &  0.1  &  0.01  &  0.04  \\ 
23  &  -0.04  &  -0.05  &  -0.05  &  0.0  \\ 
24  &  -0.05  &  0.08  &  0.01  &  0.03  \\ 
25  &  -0.0  &  0.11  &  0.05  &  0.03  \\ 
26  &  0.12  &  -0.03  &  0.04  &  0.04  \\ 
27  &  -0.0  &  0.09  &  0.04  &  0.02  \\ 
28  &  -0.01  &  0.08  &  0.03  &  0.02  \\ 
29  &  -0.06  &  0.12  &  0.03  &  0.04  \\ 
30  &  0.01  &  0.05  &  0.03  &  0.01  \\ 
31  &  -0.12  &  -0.03  &  -0.08  &  0.02  \\ 
32  &  0.02  &  0.02  &  0.02  &  0.0  \\ 
33  &  -0.01  &  0.06  &  0.02  &  0.01  \\ 
34  &  0.01  &  0.0  &  0.0  &  0.0  \\ 
35  &  0.04  &  0.09  &  0.07  &  0.01  \\ 
36  &  0.02  &  0.08  &  0.05  &  0.01  \\ 
37  &  -0.08  &  0.03  &  -0.02  &  0.03  \\ 
38  &  -0.06  &  -0.02  &  -0.04  &  0.01  \\ 
39  &  0.14  &  0.26  &  0.2  &  0.03  \\ 
40  &  -0.11  &  0.01  &  -0.05  &  0.03  \\ 
41  &  -0.02  &  0.12  &  0.05  &  0.04  \\ 
42  &  -0.02  &  0.04  &  0.01  &  0.01  \\ 
43  &  -0.02  &  0.05  &  0.01  &  0.01  \\ 
44  &  0.28  &  0.37  &  0.32  &  0.02  \\ 
HD132813  &  0.0  &  0.24  &  0.12  &  0.06  \\ 
HD175588  &  -0.05  &  0.2  &  0.07  &  0.06  \\ 
HD89758  &  -0.07  &  0.2  &  0.06  &  0.06  \\ 
HD224935  &  -0.01  &  0.26  &  0.12  &  0.06  \\ 
HD101153  &  -0.08  &  0.16  &  0.04  &  0.06  \\ 
HIP54396  &  -0.02  &  0.12  &  0.05  &  0.03  \\ 

\hline
\hline
\end{tabular}
\end{table*}

\begin{table*}
\caption{Uncertainties in the derived Mg, Si, Ca and Ti abundances from each individual line arising from typical uncertainties of $\pm$ 100 K in \teff, $\pm$0.2 dex in \logg, $\pm$0.1 dex in \feh, and $\pm$0.1 km/s in $\xi_\mathrm{micro}$. We have selected seven stars with metallicities covering the range explored in this study (metallicities are mentioned in brackets below the name of each star). The abundance uncertainty, i.e. dispersion estimated from the gaussian fit to the distribution of 50 abundance estimates (estimated using 50 sets of stellar parameters) from each line is listed in each column. The mean uncertainty based on error propagation is listed in the row after each element.    }\label{table:uncert_params}
\begin{tabular}{l c c c c c c c c}
\hline
  & Index  & 10 & 44  &  11 & 30 & 29 & 41 & 40\\
 & [Fe/H]  & -0.9 dex & -0.6 dex & -0.5 dex & -0.25 dex & 0.0 dex & 0.1 dex & 0.25 dex  \\
 & $\lambda$ ($\AA$)  & $\sigma_{line}$ & $\sigma_{line}$  &  $\sigma_{line}$ & $\sigma_{line}$ & $\sigma_{line}$ & $\sigma_{line}$ & $\sigma_{line}$\\
\hline
\multirow{3}{*}{[Mg/Fe]} & 21059.76    &  0.10   &  0.08   & 0.10  & 0.08  & 0.09  & 0.11  & 0.12  \\
&  21060.89    &  0.12   & 0.12   & 0.14   & 0.07  & 0.13  & 0.13  & 0.13  \\
&  21458.87    &  0.10   & 0.10   & 0.10   & 0.10  & 0.11  & 0.12  & 0.14 \\
\hline
$\sqrt{\sum_{\sigma_{line}^{2}}}$ / N$_{lines}$ &      &  0.04   & 0.04   & 0.07   & 0.05  & 0.07  & 0.07  & 0.08 \\
\\
\hline
\hline
\multirow{4}{*}{[Si/Fe]} & 16434.93    &  0.2   &  0.33   & 0.28  & 0.27  & 0.34  & 0.30  &  0.36 \\
& 20804.20    &  0.13   &  0.12   & 0.14  & 0.15  & 0.15  & 0.15  & 0.13  \\
& 20890.37    & 0.11    &  0.07   & 0.12  & 0.10  & 0.13  & 0.11  &  0.13 \\
&  20926.14    &  0.13   &  0.10   &  0.14  & 0.15  & 0.16  & 0.11  & 0.17  \\
\hline
$\sqrt{\sum_{\sigma_{line}^{2}}}$ / N$_{lines}$  &      &  0.07   & 0.09   & 0.09   & 0.09  & 0.11  & 0.09  & 0.11 \\
\\
\hline
\hline
\multirow{6}{*}{[Ca/Fe]} & 16150.76    &  0.14   &  0.18   & 0.12  & 0.21  & 0.16  &  0.25 & 0.09  \\
& 16155.24    &  0.10   &  0.10   & 0.08  & 0.14  & 0.12  &  0.13 & 0.06  \\
&  16157.36    &  0.14   &  0.10   & 0.09   & 0.14  & 0.10  & 0.13  & 0.12  \\
&  20962.57    &  0.09   &  0.08  & 0.07   & 0.11  & 0.08  & 0.10  & 0.10 \\
&  20972.53    &  0.12   & 0.06   & 0.10   & 0.12  & 0.11  & 0.09  & 0.09 \\
\hline
$\sqrt{\sum_{\sigma_{line}^{2}}}$ / N$_{lines}$  &      &  0.05   & 0.05   & 0.04   & 0.07  & 0.05  & 0.07  & 0.05 \\
\\
\hline
\hline
\multirow{2}{*}{[Ti/Fe]} & 16330.54    &  0.14   &  0.14   & 0.09  & 0.19  & 0.14  & 0.20  & 0.08  \\
& 21149.62    &  0.17   &  0.15  &  0.11  & 0.12  & 0.09  & 0.09  & 0.10 \\
\hline
$\sqrt{\sum_{\sigma_{line}^{2}}}$ / N$_{lines}$  &      &  0.11   & 0.10   & 0.07   & 0.11  & 0.08  & 0.11  & 0.06 \\
\\
\hline
\hline
\end{tabular}

\end{table*}

\end{appendix}

\end{document}